\documentclass[usenatbib,twocolumn]{mn2e}
\pdfoutput=1
\usepackage{natbib}
\usepackage{epsfig}
\usepackage{lscape}
\usepackage{amsmath,amsfonts,amssymb,textcomp}
\usepackage{graphicx}
\usepackage{amssymb}
\usepackage{pdflscape}
\usepackage{multirow}
\usepackage{subfigure}
\usepackage{varioref}
\usepackage{hyperref}
\hypersetup{backref,colorlinks=true,citecolor=blue} 
\usepackage{psfig}
\usepackage{times}

\newcommand{\vect}[1]{\mathbf{#1}}
\renewcommand{\d}{{\rm d}}

\labelformat{equation}{\textup{#1}}

\labelformat{Fig.}{\textup{#1}}

\date{Accepted 2013 June 18. Received 2013 May 23; in original form 2013 March 1}
\pubyear{2013}

\begin{document}
\graphicspath{{images/}}
\title[Gas rotation in galaxy clusters]{Gas rotation in galaxy clusters: signatures and detectability in X-rays} 
 \author[M. Bianconi et al.]{Matteo Bianconi$^{1,}$\thanks{matteo.bianconi@uibk.ac.at}, Stefano Ettori$^{2,3}$, Carlo Nipoti$^4$\\
 $^{1}$Institut f\"{u}r Astro und Teilchenphysik, Leopold Franzens Universit\"{a}t Innsbruck,  Technikerstra\ss e 25/8, A-6020 Innsbruck, Austria\\
 $^{2}$INAF, Osservatorio Astronomico di Bologna, via Ranzani 1, I-40127 Bologna, Italy\\
 $^{3}$INFN, Sezione di Bologna, viale Berti Pichat 6/2, I-40127 Bologna, Italy \\
 $^{4}$Dipartimento di Fisica e Astronomia, Universit\`{a} di Bologna, viale Berti-Pichat 6/2, I-40127 Bologna, Italy}

\maketitle

\begin{abstract}

We study simple models of massive galaxy clusters in which the
intracluster medium (ICM) rotates differentially in equilibrium in the
cluster gravitational potential. We obtain the X-ray surface
brightness maps, evaluating the isophote flattening due to the gas
rotation. Using a set of different rotation laws, we put
constraint on the amplitude of the rotation velocity, finding that
rotation curves with peak velocity up to $\sim 600 \;\rm km\,s^{-1}$
are consistent with the ellipticity profiles of observed clusters.
We convolve each of our models with the instrument response of the
X-ray Calorimeter Spectrometer on board the ASTRO-H to calculate the
simulated X-ray spectra at different distance from the X-ray
centre. We demonstrate that such an instrument will allow us to
measure rotation of the ICM in massive clusters, even with rotation
velocities as low as $\sim 100 \;\rm km\,s^{-1}$.
\end{abstract}


\section{Introduction}
Clusters of galaxies play a critical role in understanding the
formation and evolution of large-scale structure, and determining
cosmological parameters. Their mass is one of the most crucial
quantities to be evaluated. Strong \citep{smith05} and weak
(\citealt{mahdavi08}, \citealt{zhang08}) lensing measurements, and
Sunyaev \& Zel'dovich \citeyearpar[SZ]{sz1970} effect observations
(\citealt{morandi12}, \citealt{sereno13}) have become reliable mass
estimators.  Still, X-ray observations provide one of the fundamental
methods to recover the galaxy cluster mass. Most of these studies
include the assumption of the hydrostatic equilibrium for the
intracluster medium (ICM; \citealt{voit05}, \citealt{krav12}). This
hypothesis implies that only the thermal pressure of the hot ICM is
taken into account.  It has been claimed that non-thermal motions,
such as turbulence and rotation, could play a significant role in
supporting the ICM, in particular in the innermost region, and so
biasing the mass measurements based on the hydrostatic equilibrium
(\citealt{nagai07}, \citealt{fang09},
\citealt{humphrey12}, \citealt{valda11},  \citealt{suto13}). \citet{rasia06} evaluated that X-ray
measurements of relaxed clusters assuming hydrostatic equilibrium can
lead to underestimate the virial mass by up to $20\%$ (see also
\citealt{meneghetti10}, \citealt{rasia12}).  Thanks to the recent
improvements in hydrodynamical simulations, the expected pressure
support from random turbulent gas motion can be estimated
(\citealt{rasia06}, \citealt{nagai07}, \citealt{fang09},
\citealt{lau09}, \citealt{biffi11}).  \citet{rasia04} and
\citet{lau09} noticed that, in simulated clusters, the velocity
dispersion tensor $\sigma_{i,j}^2$ of the ICM is approximately
isotropic in the outskirts of clusters and it becomes increasingly
tangential at smaller radii, especially for the most relaxed
systems. \citet{fang09} showed that the support of the ICM from
rotational and streaming motions is comparable to the support from the
random turbulent pressure out to $\approx0.8\,r_{500}$, where
$r_{500}$ is the radius enclosing a mean overdensity of 500 in units
of the Universe critical density.  This scenario is confirmed by
\citet{lau12}, who considered non radiative simulations of
  cluster formation.

The direct observation of the velocity structure of the ICM would help
in understanding the robustness of the hydrostatic equilibrium
hypothesis and in revealing the mechanisms altering this
equilibrium. In particular, ICM rotation and turbulence can be induced
from mergers and accretion of external matter, and can trace
the cooling gas in the inner regions \citep{lau12}. The most
  direct way to measure gas motions in galaxy clusters would be via
  the broadening of the line profile of heavy ions and the shift of
  their centroids. This method has been discussed in detail in
  \citet{ino03} and by \citet{rebusco08}. The investigation of the
  imprint of ICM motions on the iron line profile requires
  high-resolution spectroscopy, which will become possible in the near
  future with the next-generation X-ray instruments, such as
  ASTRO-H\footnote{http://astro-h.isas.jaxa.jp/} \citep{astroh}.  
Only few observational constraints on ICM internal motions have been obtained with the currently available
instruments. \citet{sanders13}, using XMM-Newton Reflection Grating Spectrometer (RGS) 
spectra of about 60 among X-ray bright galaxy clusters, groups and elliptical galaxies, find 
an upper limit on the velocity broadening of $\lesssim500\;\rm km\,s^{-1}$ in about 
a third of the sample, with 5 targets with limits of $300\;\rm km\,s^{-1}$ or lower.

Another indirect signature of large-scale rotation in galaxy
  clusters is the ellipticity of the X-ray isophotes due to flattening
  of the underlying gas density distribution
  \citep[e.g.][]{brighenti96}. Clearly, if the halo deviates
  significantly from spherical symmetry, flattening of the X-ray
  isophotes is expected even in absence of rotation. In this context,
the triaxiality of the galaxy clusters is a standard scenario
confirmed both by the observations (\citealt{kawahara10}; see also
\citealt{morandi10}, \citealt{sereno13}) and by $N$-body numerical
simulations \citep{jing2002}.  However, it is not excluded that a
substantial part of the observed X-ray ellipticity is due to
rotational flattening \citep{buotecan96}. \citet{fang09} have
evaluated the isophote shapes of clusters observed with Chandra and
ROSAT, with a mean value of $\epsilon\approx0.25$.  \citet{lau12},
using the Chandra and Rosat/PSPC nearby cluster sample by
\citet{vikhl09}, observed a mean ellipticity of $\epsilon\approx0.18$
in the radial range $0.1\lesssim r/r_{500}\lesssim1$.

 The rotation of the ICM is potentially relevant also to studies
   of the thermal stability of the ICM itself. An important question
   to understand the evolution of galaxy clusters is whether the ICM
   is subject to thermal instability, so that cold clouds of gas can
   condense spontaneously far from the cluster centre
   \citep[e.g.][]{mathews78}. This problem has been investigated in
   detail under the assumption that the ICM does not rotate
   \citep[e.g.][]{malagoli87,Bal10}. However, recent work on the
   thermal stability of rotating plasmas \citep{nipoti10,nipposti} has
   shown that the onset of thermal instability can be significantly
   influenced by rotation. It follows that, not only X-ray based mass
   estimates, but also the question of the thermal stability of the
   ICM should be reconsidered if significant rotation is detected in
   the hot gas of galaxy clusters.

 In this work, we study the signatures of the presence of rotation
  of the ICM in observable properties of models representative of
  massive X-ray luminous galaxy clusters. We present simple rotation
velocity laws consistent with the observed ellipticity profiles.
Furthermore, we demonstrate the capability of the the next generation
X-ray instruments, such as the X-ray Calorimeter Spectrometer on board
the ASTRO-H satellite, in detecting and discriminating between
different type of motions in the ICM distribution. This calorimeter,
thanks to its excellent energy resolution (with a requirement of
$\Delta E\approx7\;\rm eV$ and a goal of $4\;\rm eV$), will allow us
to directly detect non-thermal contributions to the cluster pressure
support, such as measuring the line centroid Doppler shift due to the
rotational velocity in the ICM and the emission line broadening due to
turbulent motions.
\section{The cluster models}

\subsection{The fluid equations}
\label{sec:fluideq}

Here we present the fluid equations governing our models of rotating
ICM. We consider the equations describing an axisymmetric
gaseous system in permanent rotation in equilibrium in an
axisymmetric gravitational potential $\Phi$. The imposed symmetry
implies that the physical variables depend only on the cylindrical
coordinates $R$ and $z$. The stationary hydrodynamic equations
governing the gas distribution are then
\begin{equation}
 \begin{cases}
\dfrac{1}{\rho}\dfrac{\partial P}{\partial z}=-\dfrac{\partial \Phi}{\partial z}\\
\dfrac{1}{\rho}\dfrac{\partial P}{\partial R}=-\dfrac{\partial \Phi}{\partial R} + \Omega^2 R,\label{sys}
\end{cases}
\end{equation}
where $\rho$, $P$ and $\Omega$ denote the gas density, pressure, and
angular velocity, respectively. The gas rotational velocity is given
by $u_{\varphi}=\Omega R$, whilst we assume $u_R=u_z=0$; $\Phi$
represents the total gravitational potential, including the stars, gas
and dark matter contribution.

In general $\Omega=\Omega(R,z)$, but for simplicity we consider here
the case of cylindrical rotation $\Omega=\Omega(R)$.  The
\textit{Poincar\'{e}-Wavre} theorem (\citealt{tassoul1978}) states
that cylindrical rotation (i.e. $\Omega$ depends only on $R$) is a
necessary and sufficient condition to have a barotropic
stratification, in which the isobaric and isodensity surfaces
coincide, so $P=P(\rho)$. When $\Omega=\Omega(R)$, introducing the
effective potential
\begin{equation}
 \Phi_{\rm eff}(R,z)= \Phi(R,z) - \int^R\Omega^2(R') R' dR',
\end{equation}
the system of
equations~(\ref{sys}) can be written as
\begin{equation}
 \dfrac{\nabla P}{\rho}= -\nabla\Phi_{\rm eff}, \label{eqpoi} 
\end{equation}
which shows that a barotropic fluid can be seen as in hydrostatic
equilibrium in the effective potential.
Assuming that the distribution is polytropic, so
  $P/P_0=(\rho/\rho_0)^{\tilde\gamma}$, in which $P_0=P(\vect{x}_0)$
  and $\rho_0=\rho(\vect{x}_0)$, where $\vect{x}_0$ is a
  reference point, and $\tilde\gamma$ is the polytropic index, equation~(\ref{eqpoi}) becomes
\begin{equation}
 \tilde\gamma \dfrac{k_{\rm B} T_0}{\mu m_{\rm p}}\frac{\rho^{\tilde\gamma-1}}{\rho_0^{\tilde\gamma-1}}\nabla \rho = -\rho \nabla \Phi_{\rm eff}\label{eq4},
\end{equation}
where $T_0=T({\bf x}_0)$.
As $\Phi_{\rm eff}=\Phi_{\rm eff}(\rho)$, we have $\nabla\Phi_{\rm
  eff}=(\d\Phi_{\rm eff}/\d\rho)\nabla\rho$, leading to
\begin{equation}
 \int^{\rho(\vect{x})/\rho_0}_1 \tilde\gamma \dfrac{k_{\rm B} T_0}{\mu m_{\rm p}}\rho'^{\tilde\gamma-2} \d\rho'= -\int^{\Phi_{\rm eff}(\vect{x})}_{\Phi_{\rm eff,0}} \d\Phi'_{\rm eff}\label{state},
\end{equation}
where $\rho'=\rho/\rho_0$ and $\Phi_{\rm eff,0}=\Phi_{\rm
    eff}(\vect{x}_0)$. When
  $\tilde\gamma\neq 1$ we can write
\begin{equation}
 \rho(\Phi_{\rm eff})= \rho_0 \biggl[1+\dfrac{\tilde\gamma-1}{\tilde\gamma}\dfrac{\mu m_{\rm p}}{k_{\rm B} T_0}(\Phi_{\rm eff,0}-\Phi_{\rm eff})\biggr]^{\frac{1}{\tilde\gamma-1}}.\label{poli}
\end{equation}
 When the distribution is isothermal ($\tilde\gamma=1$, $T=T_0$)
  equation~(\ref{eq4}) becomes
\begin{equation*}
\dfrac{k_{\rm B} T_0}{\mu m_{\rm p}}\frac{\nabla\rho}{\rho}= -\nabla\Phi_{\rm eff} ,
 \end{equation*}
 leading to the density distribution
\begin{equation}
\rho(R,z)\equiv \rho_{\rm 0} e^{-[\Phi_{\rm eff}(R,z)-\Phi_{\rm eff,0}](\mu m_{\rm p}/k_{\rm B} T_0)}. \label{ro_iso}
\end{equation}

\subsection{The mass distribution}\label{r200}

One of the ingredients of our cluster models is the total
gravitational potential $\Phi$, which is determined by the cluster
mass distribution.  We do not attempt to model a particular cluster,
but we present an idealised case study representative of massive
clusters. For our case study, we consider a model cluster with a total
mass fixed to $ M_{200}=10^{15}\; M_{\odot}$, where $ M_{200}$ is the
mass measured at the radius $r_{200}$, within which the mean cluster
density is 200 times the cosmic critical density at the cluster
redshift. The $N$-body simulations of the structure formation in
  cold dark matter models show that the density distribution of the
  dark matter haloes can be well characterized, for given $ M_{200}$,
  with only two parameters, the scale radius $r_{\rm s}$ and the
concentration parameter $c_{200}$ (\citealt{navarro1996}; hereafter
NFW), which are related to $r_{200}$ by $r_{200}=c_{200}r_{\rm s}$. In
order to fix $c_{200}$, we use the observational $c_{200}-M_{200}$
relation from
\citet{ettori2010} that follows the original parametrization introduced from \citet{dolag04}:
\begin{equation}
 \log_{10}[c_{200}\times(1+z)]= A +B\log_{10}\left(\dfrac{M_{200}}{10^{15}M_{ \odot}}\right)\label{cm},
\end{equation}
where $A\simeq0.6$ and $B\simeq-0.4$. This relation points out the
tendency of more massive cluster haloes to be less concentrated than
the smaller haloes.  For $ M_{200}=10^{15} M_{\odot}$, we get
$c_{200}\simeq3.98$, which we assume for our models. We can now
evaluate $r_{200}$ (and then $r_{\rm s}$) of the cluster model from
the equation
\begin{equation}
 M_{200}= 200\rho_{\rm crit} \dfrac{4}{3}\pi r_{200}^3,
\end{equation}
where $\rho_{\rm crit}=3H(z)^2/8\pi G$ is the critical density at the cluster
redshift $z$ and $H(z) = H(0)\times \left[\Omega_{ \Lambda} +
  \Omega_{\rm m}(1 + z)^3\right]^{1/2}$ is the Hubble parameter. We consider
$z=0$, $H(0)=70\;\rm km\,s^{-1}, \Omega_{\Lambda}=0.73$ and $\Omega_{\rm m}=0.27$.
Thus, $r_{200}\simeq 2066\rm\; kpc$ and $r_{\rm s}\simeq519\rm\; kpc$,
  which we adopt in our model cluster. We assume also that the mass
distribution of the cluster is spherical with NFW
density profile. 
The total gravitational potential is
\begin{equation}
 \Phi(r) =  -4\pi G\rho_{\rm s} r_{\rm s}^2\frac{{\rm ln}(1+r/r_{\rm s})}{r/r_{\rm s}},
\end{equation}
where
\begin{equation}
 \rho_{\rm s}=\dfrac{200}{3}\dfrac{\rho_{\rm crit}c_{200}^3}{
\ln(1+c_{200})-c_{200}/(1+c_{200})}.
\end{equation}
Therefore, we do
not include explicitly the self gravity of the ICM (which is not
spherically symmetric because of rotation): this is justified to first
order because dark matter is known to be dominant. 

The assumption of a spherical dark matter halo deserves some discussion. 
Of course, it would be possible to build analogous models with ellipsoidal 
dark matter halos (see, e.g., \citealt{rodi12} and \citealt{bhab}), 
which are expected to be more realistic. The deviation from spherical symmetry of 
the dark matter halo is especially relevant in the present context because it can 
contribute  to the flattening of the ICM. However, for this very reason, we assume 
here spherically symmetric dark matter halo to disentangle the effect of rotation 
from the effect of  a flattened gravitational potential. Of course, with this 
choice we maximize the role of rotational flattening, so the velocity profiles 
here obtained must be considered upper limits to the line-of-sight rotational 
velocity, consistent with the observed isophote flattening.

\subsection{The rotations laws}\label{vlaws}

\begin{figure}
  \includegraphics[width=1.02\linewidth, keepaspectratio]{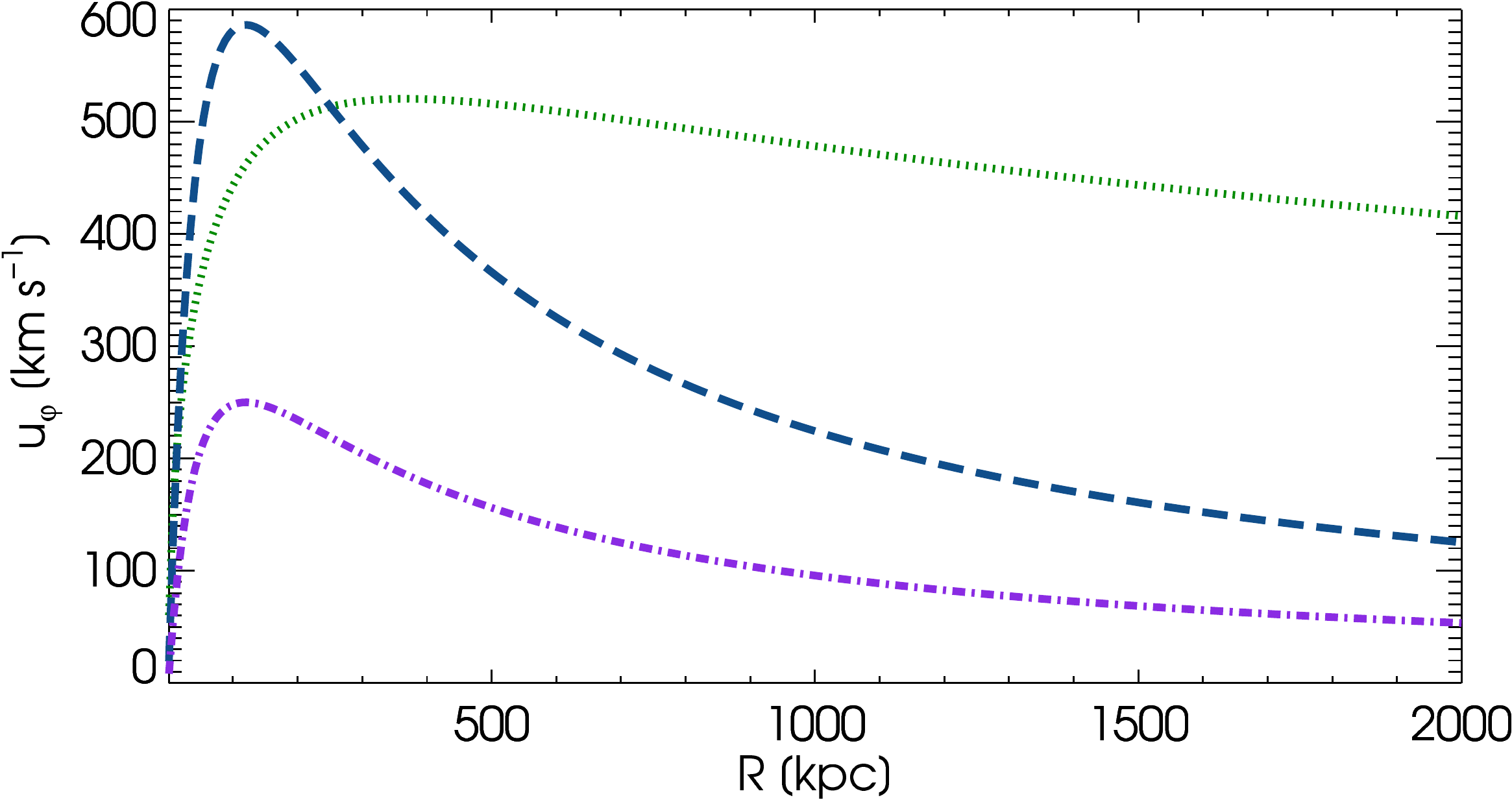}
\caption{The velocity profiles VP1 (dotted green line), VP2 (dashed
  blu line) and VP3 (dot-dashed purple line) defined in
  Section~\ref{sec:rot}.}\label{vpro}
 \end{figure}

In our models we assume that the ICM rotates differentially with
  rotation speed constant on cylinders, so $u_\varphi=u_\varphi(R)$.  The
  ICM velocity pattern is currently almost unconstrained, both
  observationally and theoretically. This allows us to choose the
  velocity profile without particular restraint. Here we consider the
following rotation velocity laws:

\begin{equation}
 u_{ \varphi}^2= u_0^2\left[\dfrac{\ln(1+S)}{S}-\dfrac{1}{S+1}\right], \label{eq:rot1}
\end{equation}
\begin{equation}
 u_{ \varphi}^2= u_0^2\dfrac{S^2}{(1+S)^4},  \label{eq:rot2}
\end{equation}
where $S=R/R_0$, and $R_0$ is the characteristic radius of the velocity law.
 Equation~(\ref{eq:rot1}) mimics the circular velocity profile of
  the NFW model, with a null value in the centre, followed by a steep
rise at intermediate radii and a shallow decrease in the outer
regions. Equation (\ref{eq:rot2}) presents a steeper increase at small
radii followed by a stronger decrease in the outer region. Based on
the above rotation laws, we choose three specific velocity profiles:
\begin{itemize}
\item[VP1:]
equation (\ref{eq:rot1}) with $u_0=1120\, \rm km\;s^{-1}$ and $R_{
  0}=170\;\rm kpc$; 
\item[VP2:] equation (\ref{eq:rot2}) with $u_0=2345\;
\rm km\;s^{-1}$ and $R_{0}=120\,\rm kpc$;
\item[VP3:]
equation (\ref{eq:rot2}) with $u_0=1000\, \rm km\; s^{-1}$ and $R_{
  0}=120\;\rm kpc$. 
\end{itemize}
 The choice of the values of parameters is such that these profiles
 lead to ICM ellipticity (described in Section~\ref{500}) comparable with those measured in observed
 clusters. The velocity profiles VP1, VP2 and VP3 are displayed in Fig.~\ref{vpro}.

\label{sec:rot}
\begin{table}
 \begin{center}
  \begin{tabular} {c c}
\hline
\hline
Parameter& Value\\
\hline
\hline
$M_{200}$ & $10^{15} \; M_{ \odot}$\\
$M_{\rm ICM}$ & $ 1.3\times10^{14}\; M_{ \odot}$\\
$r_{\rm s}$& $ 519 \;\rm kpc$\\
$r_{200}$ & $ 2066 \;\rm kpc$\\
$c_{200}$&$ 3.98$\\
$Z$ & $0.3\; Z_{\rm \odot}$\\
\hline
\hline
  \end{tabular}
 \end{center}
\caption{Values of the parameters common to all models.}\label{common}
\end{table}
\begin{table}
  \begin{center}
\begin{tabular}{c c c c c}
\hline
\hline
Model & $\tilde{\gamma}$ & $n_{c}\;\rm (10^{-3}\,cm^{-3})$ & Rotation pattern  \\
\hline
\hline 
I1& 1 & $4.92$& VP1 \\ \hline
I2& 1 & $5.24$& VP2 \\ \hline
I3& 1 & $9.33$& VP3  \\ \hline
NI1& 1.14 & $9.88$&  VP1\\ \hline
NI2& 1.14 & $10.23$&  VP2\\ \hline
NI3& 1.14  & $17.90$&  VP3\\ \hline
\hline
 \end{tabular}
  \end{center}
 \caption{ Specific properties of the models introduced in Section~\ref{mod_pre}. Name of the model
     (Column 1), polytropic index $\tilde\gamma$ (Column 2), gas central number
     density $n_{c}$ (Column 3), and name of the rotation pattern, as
     given in Section~\ref{sec:rot} (Column 4).}\label{table2}
 \end{table}

\subsection{The gas fraction and temperature}\label{mod_pre}
The gas fraction $f_{\rm gas}=M_{\rm ICM}/M_{\rm 200}$, where
  $M_{\rm ICM}$ is the total mass of the ICM, can be estimated using
the observational relation \citep{eckert2011}
\begin{equation}
 f_{\rm gas} (<r_{200}) = (0.15 \pm 0.01) \left(\frac{k_{\rm B}T}{10 \rm \,keV}\right)^{0.478}\label{eke},
\end{equation}
 \begin{figure}
 \centering
  \includegraphics[width=\linewidth, keepaspectratio]{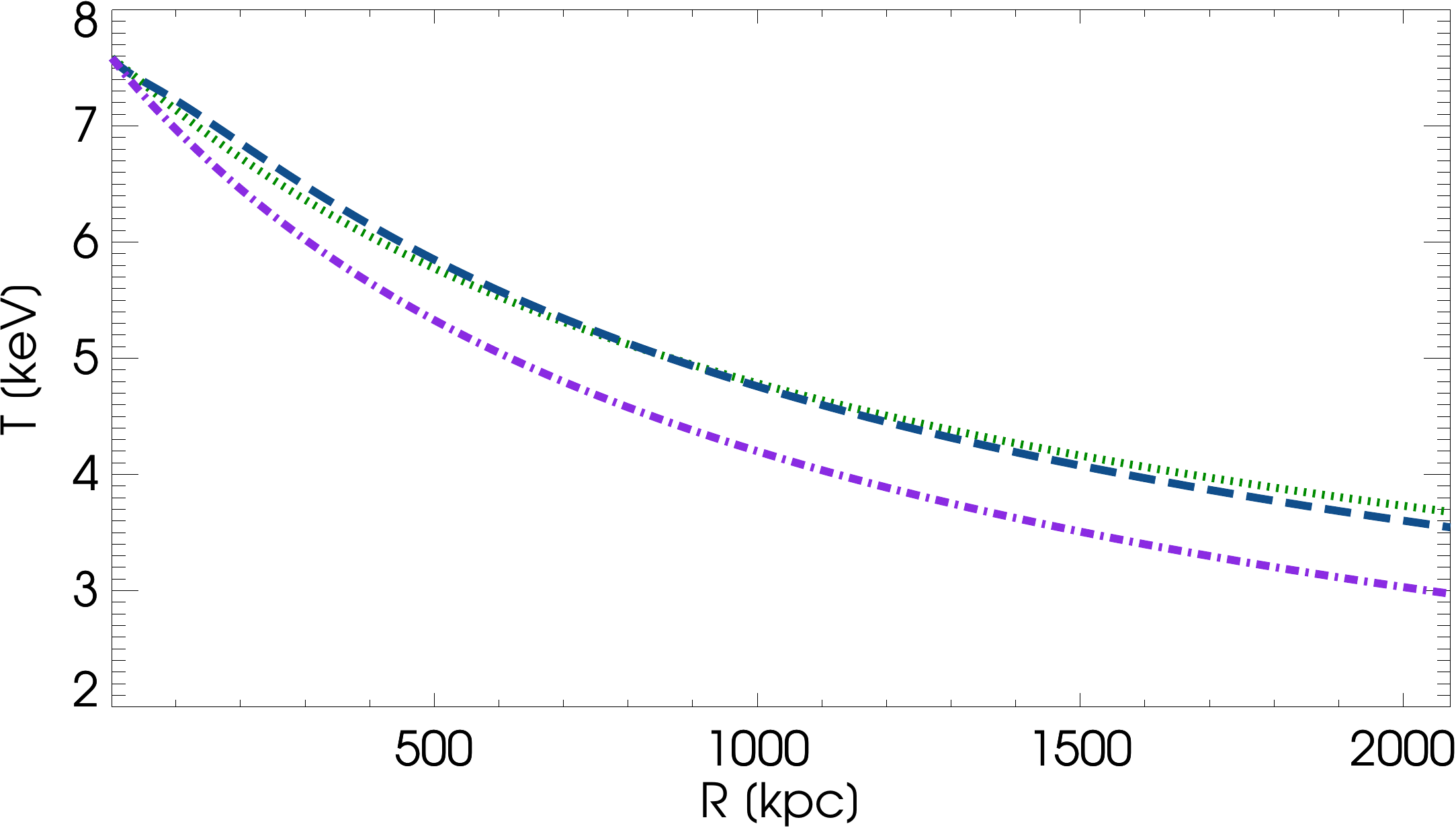}
\caption{The temperature profiles along $z=0$ for the non-isothermal
  models NI1 (dotted green line),  NI2 (dashed blu line)
  and  NI3 (dot-dashed purple line).}\label{tprof}
 \end{figure}
where $T$ is the gas temperature. In the contest of the
self-similar model of galaxy clusters, $T$ is expected to be
related to the cluster mass by $M_{200}\propto T^{3/2}$.
This expectation is confirmed by the observations of massive clusters
for which \citep{arnaud2005}
\begin{equation}
 h(z)M_{200}=A_{200}\left(\dfrac{k_{\rm B} T}{5\,\rm keV}\right)^{\alpha}, \label{arnaud}
\end{equation}
where $A_{200}\simeq5.74 \times 10^{14}\, M_{ \odot}\,$ and
$\alpha\simeq1.49$. For our models with $ M_{200}=10^{15}\;
 M_{\odot}$ at $z=0.1$, through equation~(\ref{arnaud}) we obtain the ICM
   temperature $T\simeq 8.7\times10^{7}\;\rm K$
   leading (via equation~\ref{eke}) to the gas fraction $f_{\rm gas}\simeq0.13$, which we adopt
   for our case study. This value of $f_{\rm gas}$ is in good agreement
   with \citet{maughan2008} and \citet{vikhlinin2006}.

 We consider two classes of models: isothermal models with
  polytropic index $\tilde\gamma=1$, and non-isothermal models with
  $\tilde\gamma=1.14$ (see Section~\ref{sec:fluideq}), in agreement
  with the observational constraints by \citet{markevitch1998} and \citet{vikhlinin2006}.
 In Section~\ref{sec:obs} we present results for three rotating isothermal models (I1, I2 and I3) and
 three rotating non-isothermal models (NI1, NI2 and NI3), characterized by the rotation laws 
  VP1, VP2 and VP3, respectively (see Section~\ref{sec:rot}). In all cases we assume that 
the central temperature is $T_c = 8.7 \times 10^7\, K$.
  When $\Phi$, $f_{gas}$,$\tilde\gamma$, $u_{\phi}$ and $T_{c}$ are fixed, the temperature 
 and density distributions can be computed from equation~(\ref{poli}), (\ref{ro_iso}),
 and $T/T_0=(\rho/\rho_0)^{\tilde\gamma-1}$. While, by construction,
 in the isothermal models the temperature is independent of position, in the non-isothermal models the temperature
 decreases for increasing distance from the centre. The temperature profiles in the $z=0$ plane 
  of the three non-isothermal models are plotted in Fig.~\ref{tprof}.

  In summary, the parameters common to all models are listed in
  Table~\ref{common}, while the specific values of the parameters for
  each model, including the gas central number density $n_{c}$, are
  given in Table~\ref{table2}.

\section{Observables}\label{sec:obs}

We describe here observable quantities of our rotating ICM models
  to be compared with current or forthcoming observations of real
  galaxy clusters. We consider X-ray surface brightness maps 
  and X-ray spectra, both of which are affected by rotation.
We note that the models here presented are not 
representative of cool core clusters. We defer the discussion  of the effect of cool cores
to Section~\ref{sec:cool}.

\subsection{X-ray maps and surface brightness profiles}

 \begin{figure*}
  \includegraphics[width=0.33\linewidth, keepaspectratio]{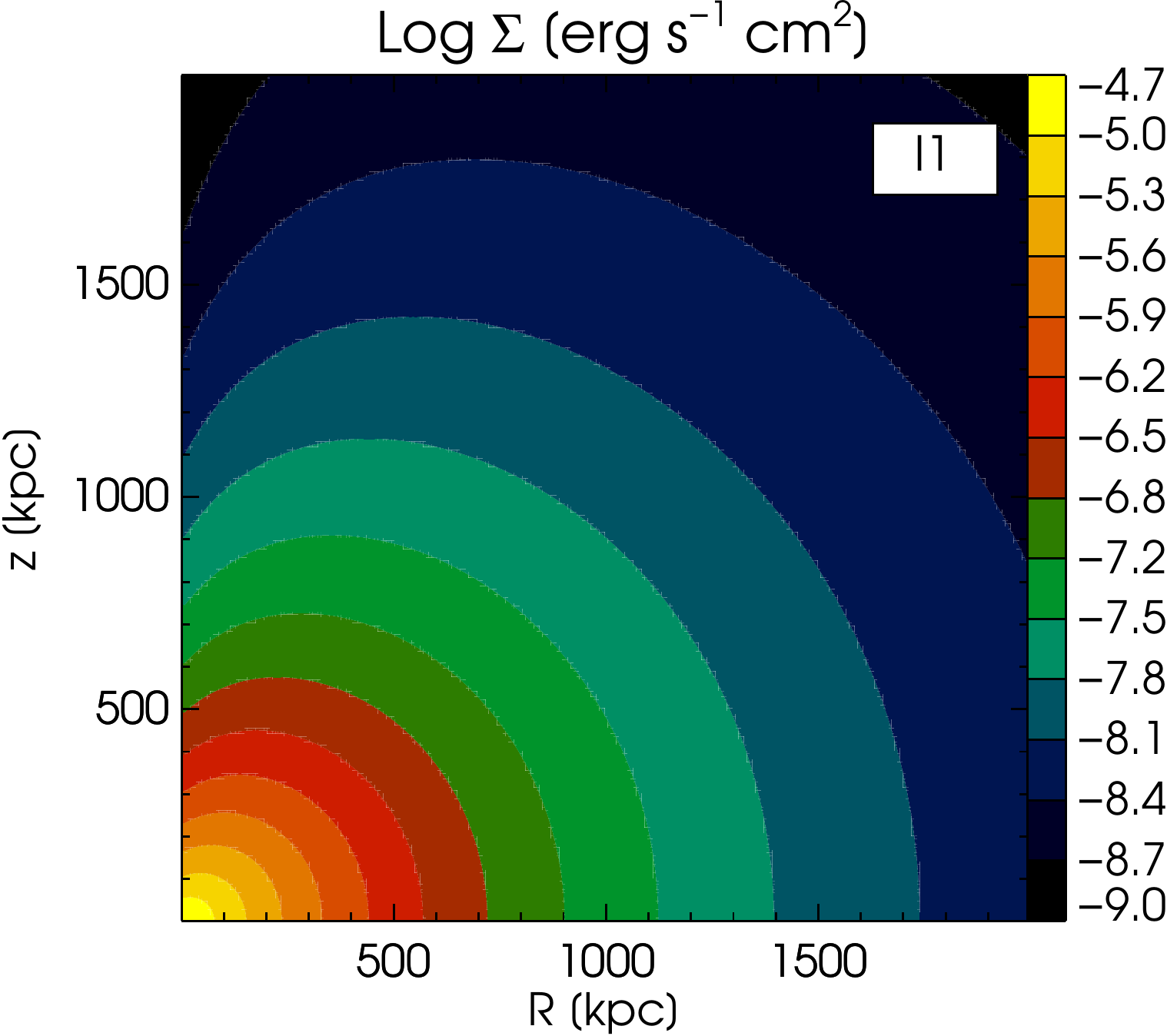}\includegraphics[width=0.33\linewidth, keepaspectratio]{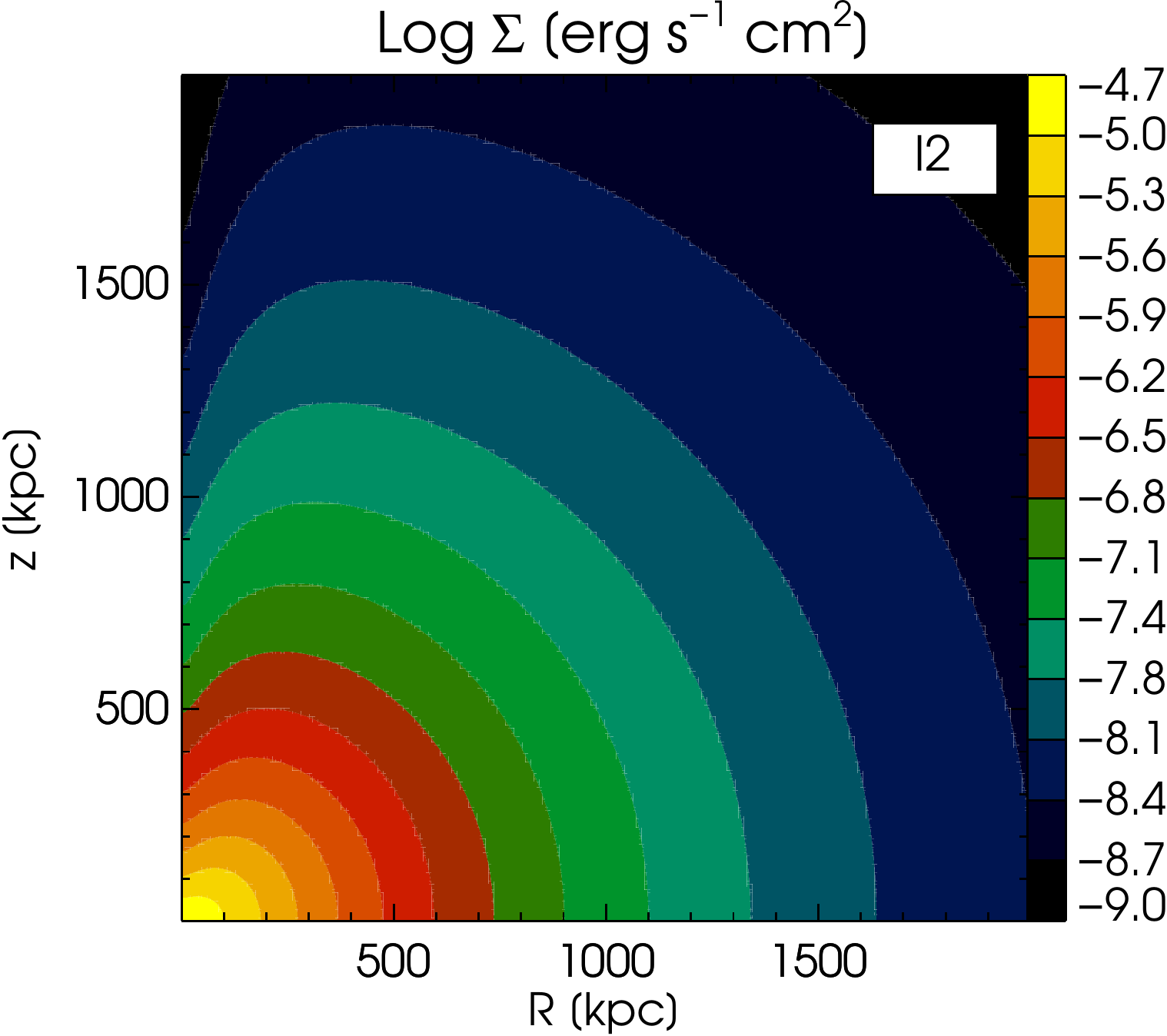}\includegraphics[width=0.33\linewidth, keepaspectratio]{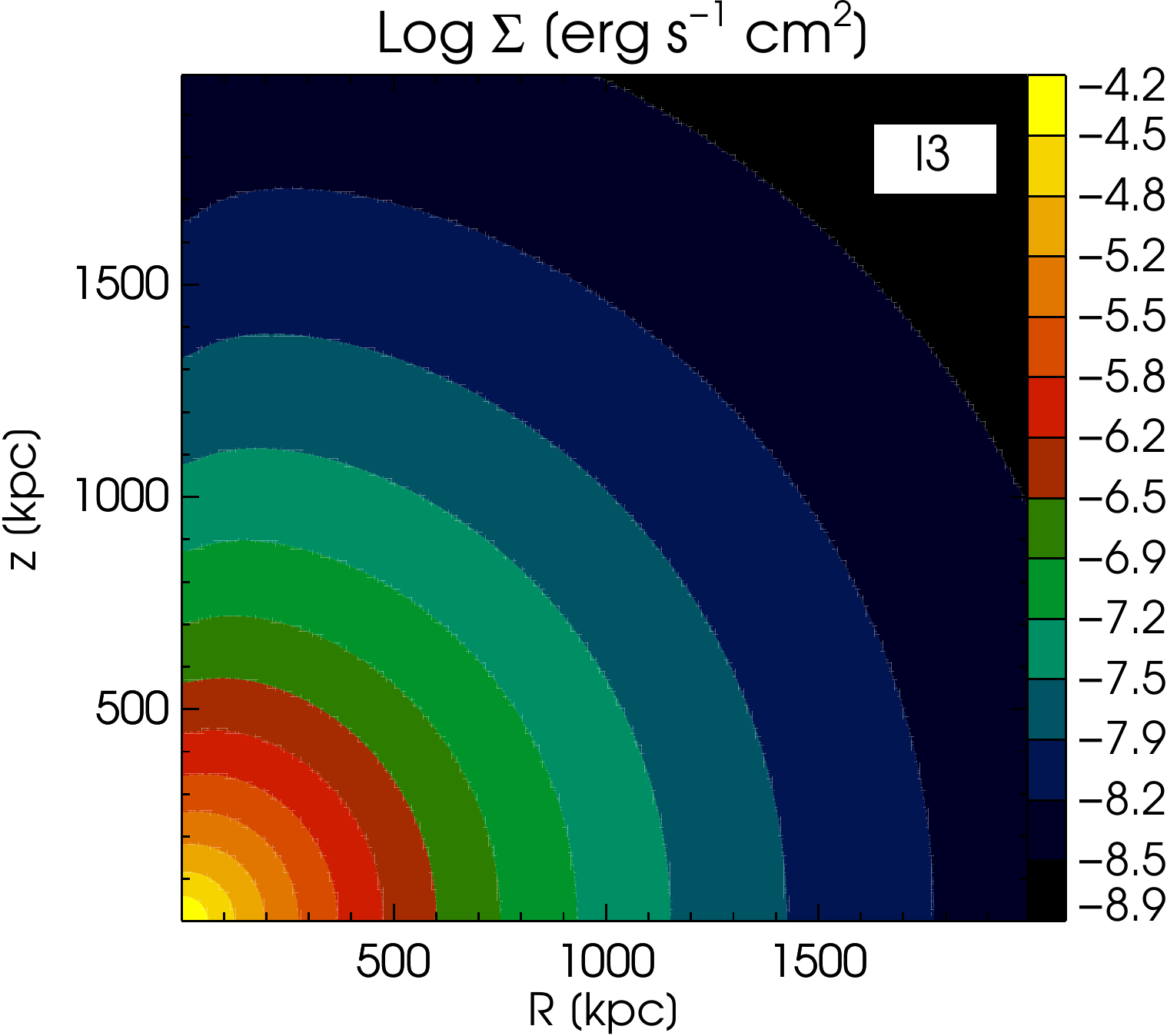}
  \includegraphics[width=0.33\linewidth, keepaspectratio]{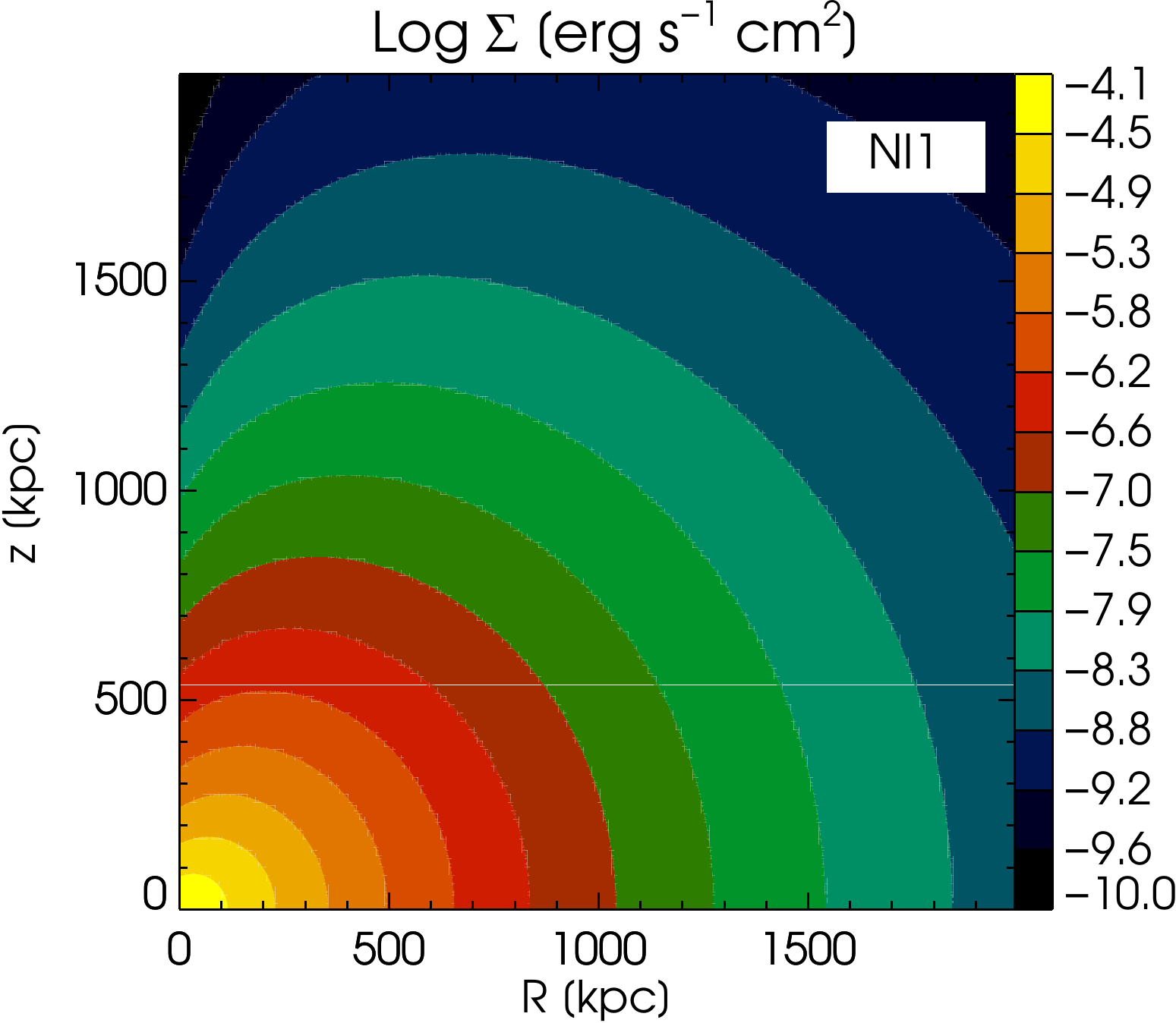}\includegraphics[width=0.33\linewidth, keepaspectratio]{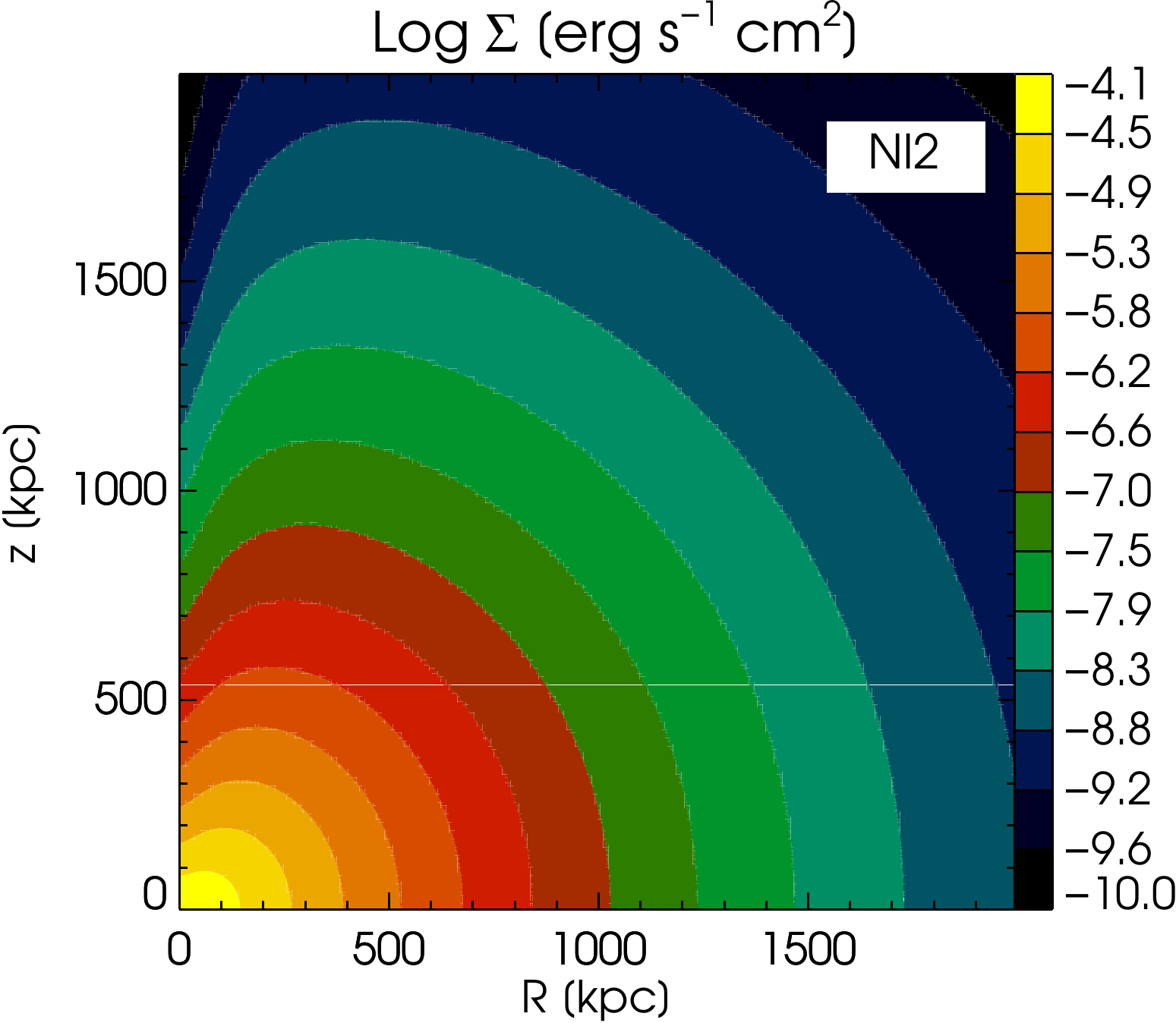}\includegraphics[width=0.33\linewidth, keepaspectratio]{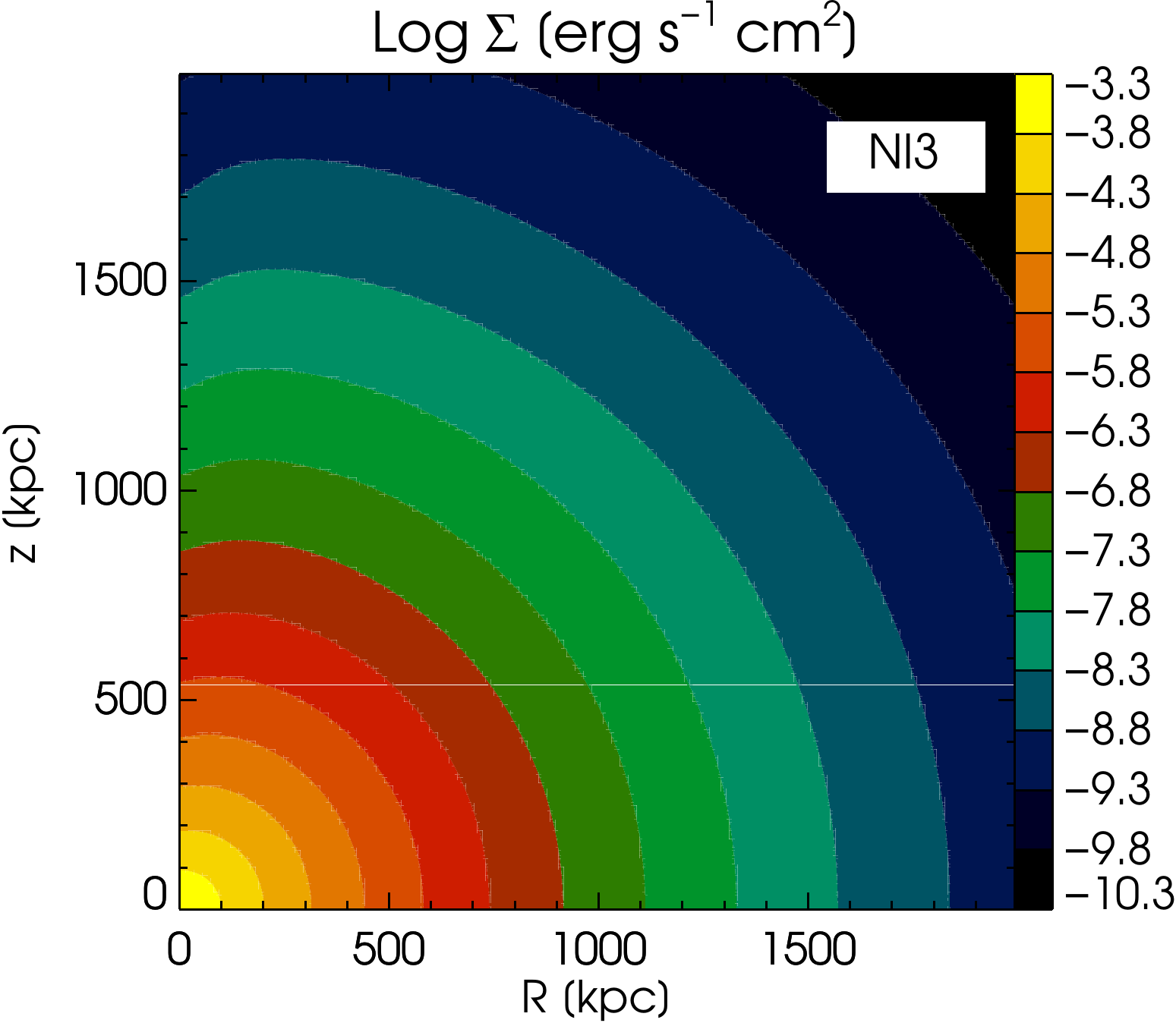}
\caption{The surface brightness maps of the isothermal (upper panels) and non-isothermal (lower panels) rotating models.}\label{iso}
 \end{figure*}

\begin{figure*}
\centering
\hbox{
\includegraphics[width=0.48\linewidth, keepaspectratio]{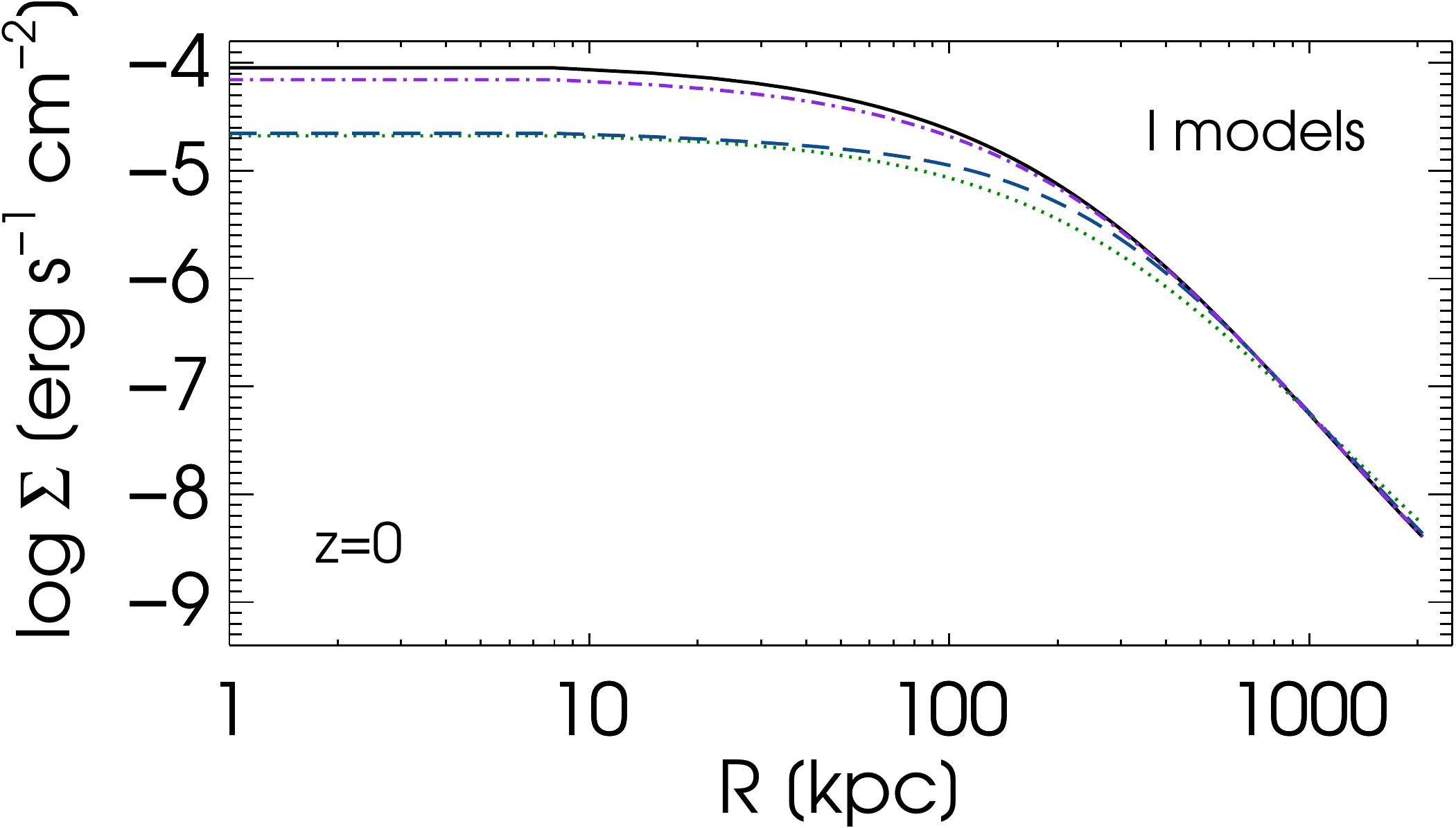}\includegraphics[width=0.48\linewidth, keepaspectratio]{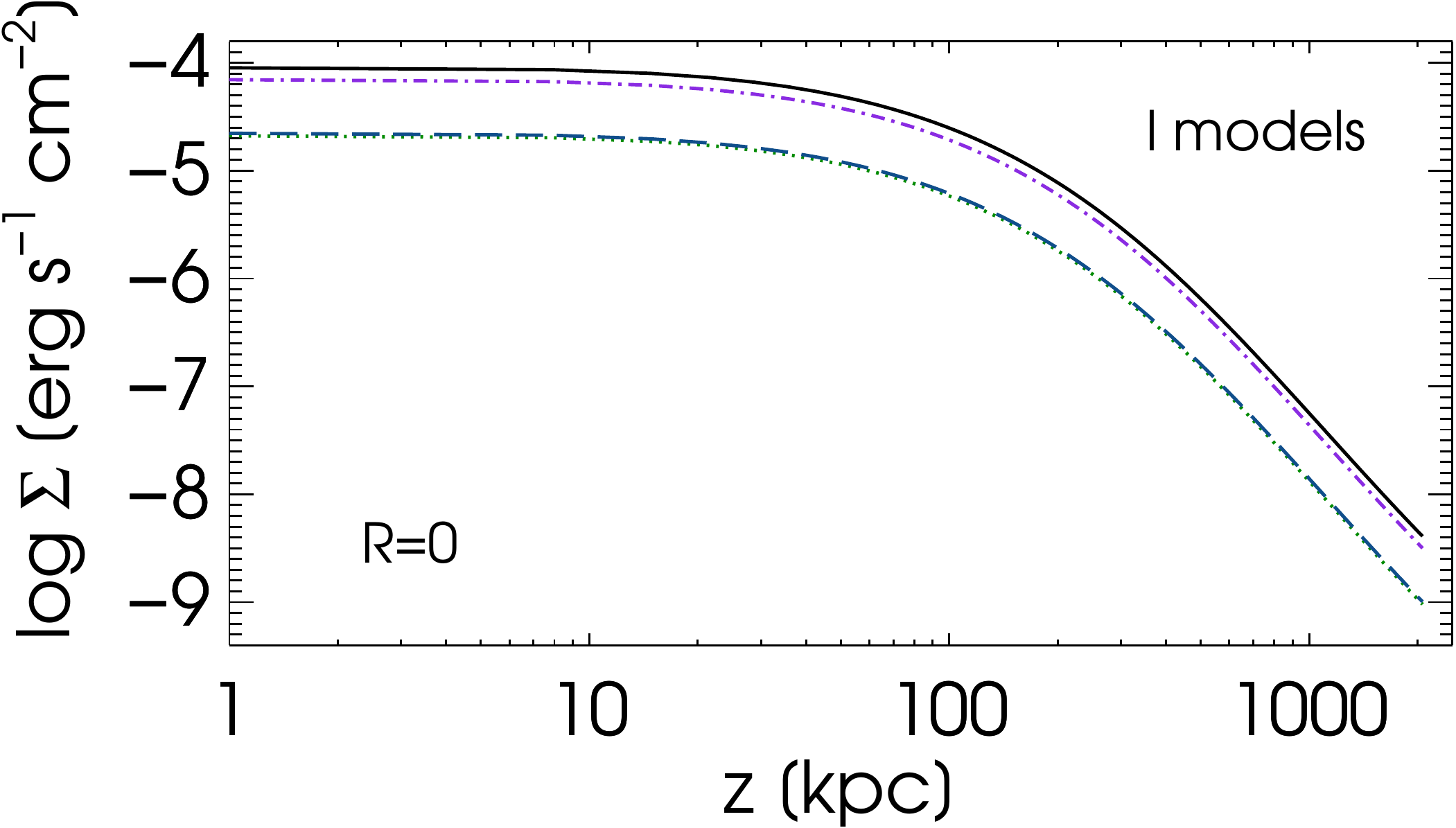}}
\hbox{
\includegraphics[width=0.48\linewidth, keepaspectratio]{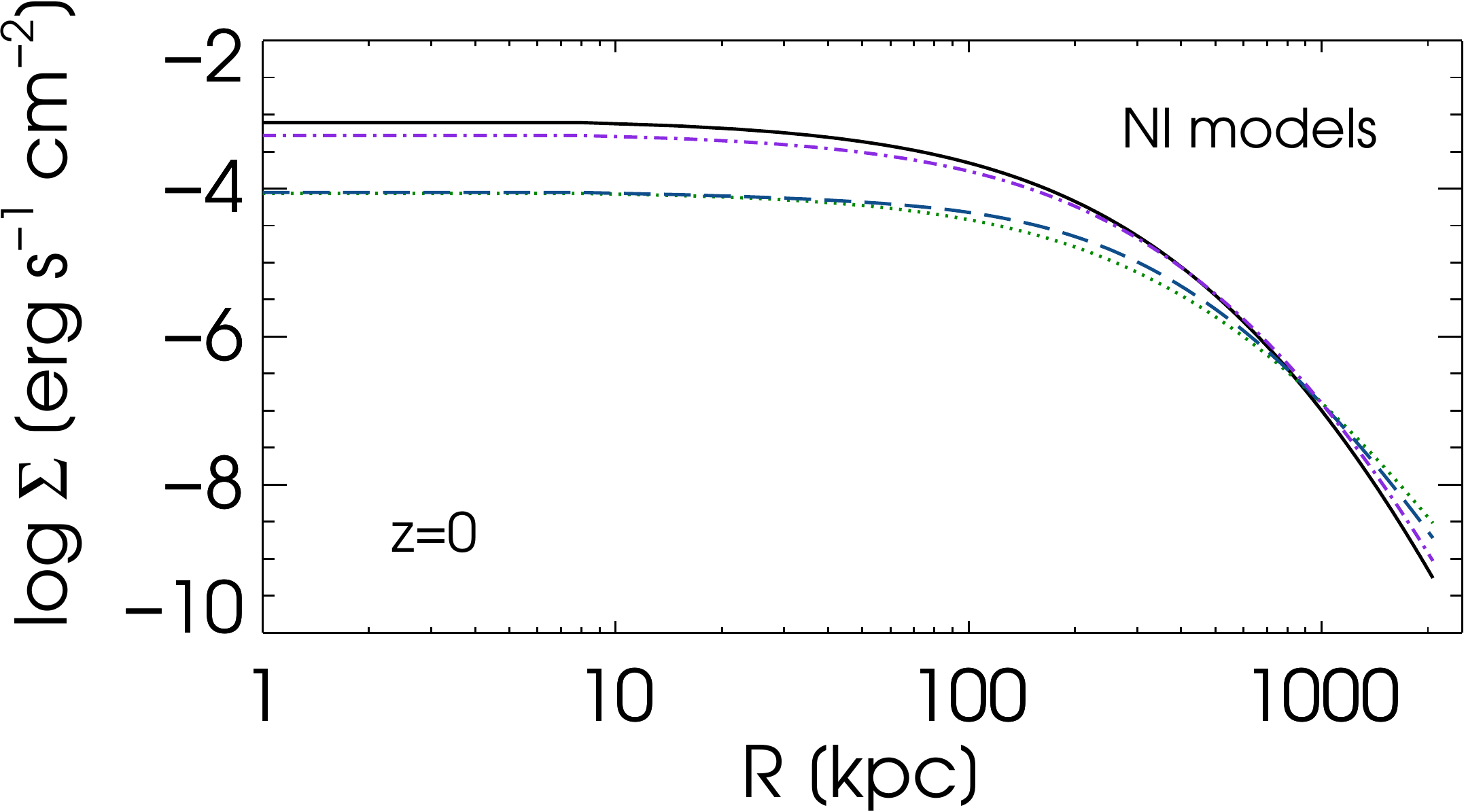}\includegraphics[width=0.48\linewidth, keepaspectratio]{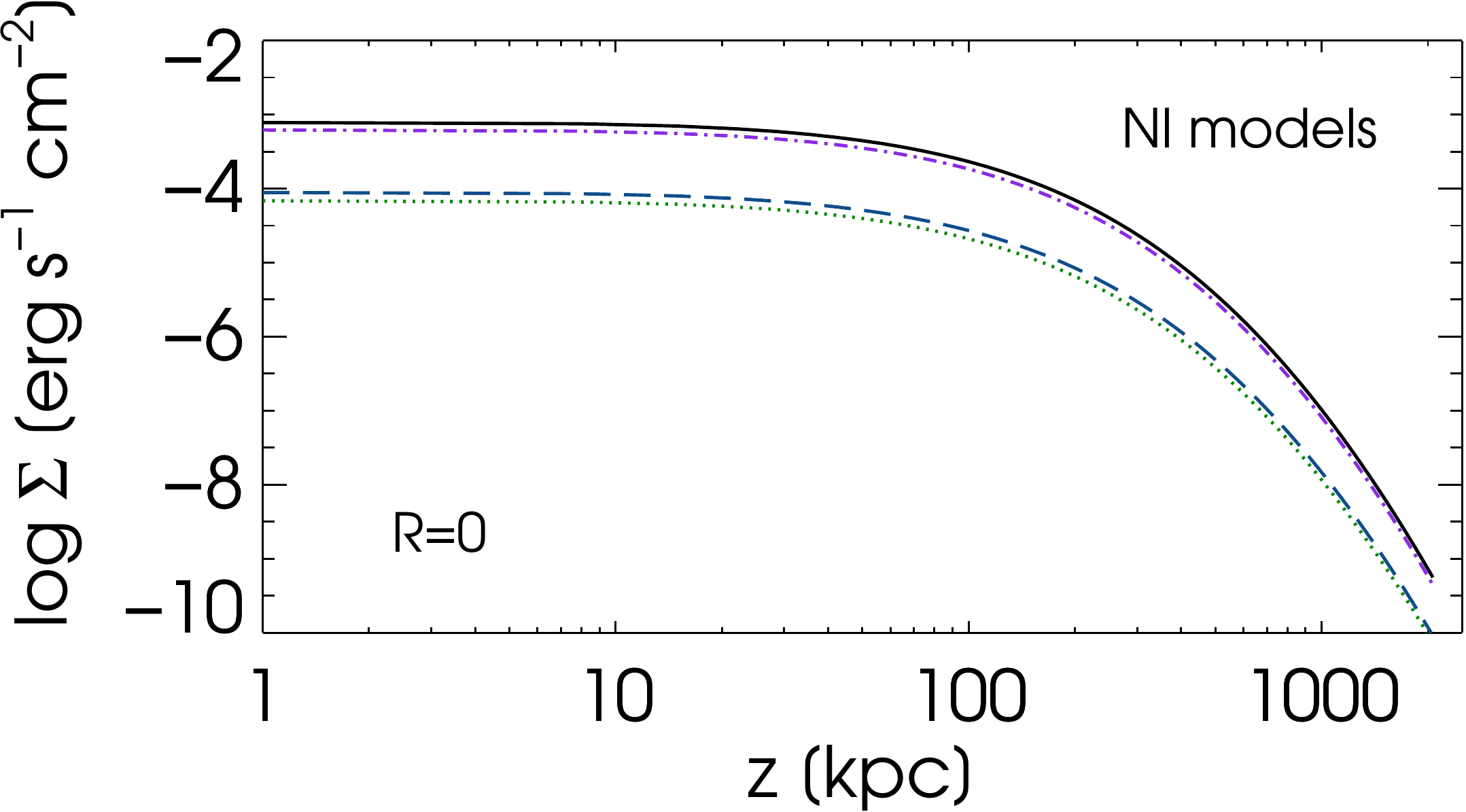}}
\caption{The surface brightness profiles along $z=0$ (left panels) and
  $R=0$ (right panels), for the isothermal (upper panels) and
  non-isothermal (lower panels) models. Specifically, the I1 and
    NI1 (dotted green line), I2 and NI2 (dashed blu line), and I3 and
    NI3 (dot-dashed purple line) models are shown. The solid lines
  represent the corresponding non-rotating models.}\label{brit_pro}
 \end{figure*}
 For each model, the surface brightness map is produced to perform a direct comparison with
the observations. The amount of rotational motion directly affects the
ICM shape, making the isophotes deviate from circular symmetry.
In order to estimate the maximum effect of rotational flattening, we
assume that our models are seen edge-on. In this hypothesis, the
surface brightness is
\begin{equation}
 \Sigma(R,z)=2\int^{\infty}_{R}\dfrac{\dot{E}(R',z)R'dR'}{\sqrt{R'^2-R^2}},
\end{equation}
where $\dot{E}=n_{\rm e}n_{\rm
  p}\Lambda(T)$ is the gas cooling rate. Here $n_{\rm e}$ and $n_{\rm p}$ are, respectively,
the electron and proton number densities, and $\Lambda(T)$ is the
cooling function.  In particular, we adopt the cooling function by
\citet{tozzi2001}
\begin{equation}
 \Lambda=C_1(k_{\rm B} T)^{\alpha}+ C_2(k_{\rm B} T)^{\beta} + C_3,
\end{equation}
where the exponents take the values $\alpha={-1.7}$
$\beta = 0.5$ and $C_1$, $C_2$, $C_3$ are constants depending on the the ICM metallicity: 
in all our models we assume a metallicity of $Z = 0.3 Z_{\odot}$, leading to
$C_1=8.6\times10^{-3}\,\rm erg\,cm^3\,s^{-1}\,keV^{-\alpha}$,
$C_2=5.8\times10^{-2}\,\rm erg\,cm^3\,s^{-1}\,keV^{-\beta}$,
$C_3=6.3\times10^{-2}\,\rm erg\,cm^3\,s^{-1}$.  

The resultant ICM morphology of the three rotating isothermal models
I1, I2 and I3 and of the non-isothermal models NI1, NI2 and NI3 can be
seen in Fig.~\ref{iso}. The effect of the rotation, visible as 
the flattening of the isophotes, reflects also in the surface brightness profiles along the $z=0$ and $R=0$ axes, shown
in Fig.~\ref{brit_pro}. In this figure, for a better comparison, we included the
 isothermal and non-isothermal non-rotating models, 
having the same gas fraction and central temperature as the corresponding rotating models.
 The surface brightness profiles of the rotating models along the axis 
of rotation ($R=0$) show the depletion of the inner regions, while 
the surface brightness along $z=0$ is systematically 
lower than that of the corresponding non-rotating models up to the virial radius.

From the surface-brightness maps shown in Fig.~\ref{iso} it is apparent 
that the gas distribution in our models tends to be peanut-shaped, 
with a depletion of gas along the vertical axis. This effect is particularly 
strong, because for simplicity we are assuming cylindrical 
rotation, and we expect it to be mitigated in more realistic baroclinic 
models in which a vertical gradient in the azimuthal velocity is allowed. 
 In any case, such peanut-shaped distributions would be hardly detectable 
in observed clusters, in which  the decrease of surface brightness in the 
central regions is interpreted as due to the presence of cavities in the ICM.
\begin{figure*}
\centering
\hbox{
  \includegraphics[width=0.5\linewidth, keepaspectratio]{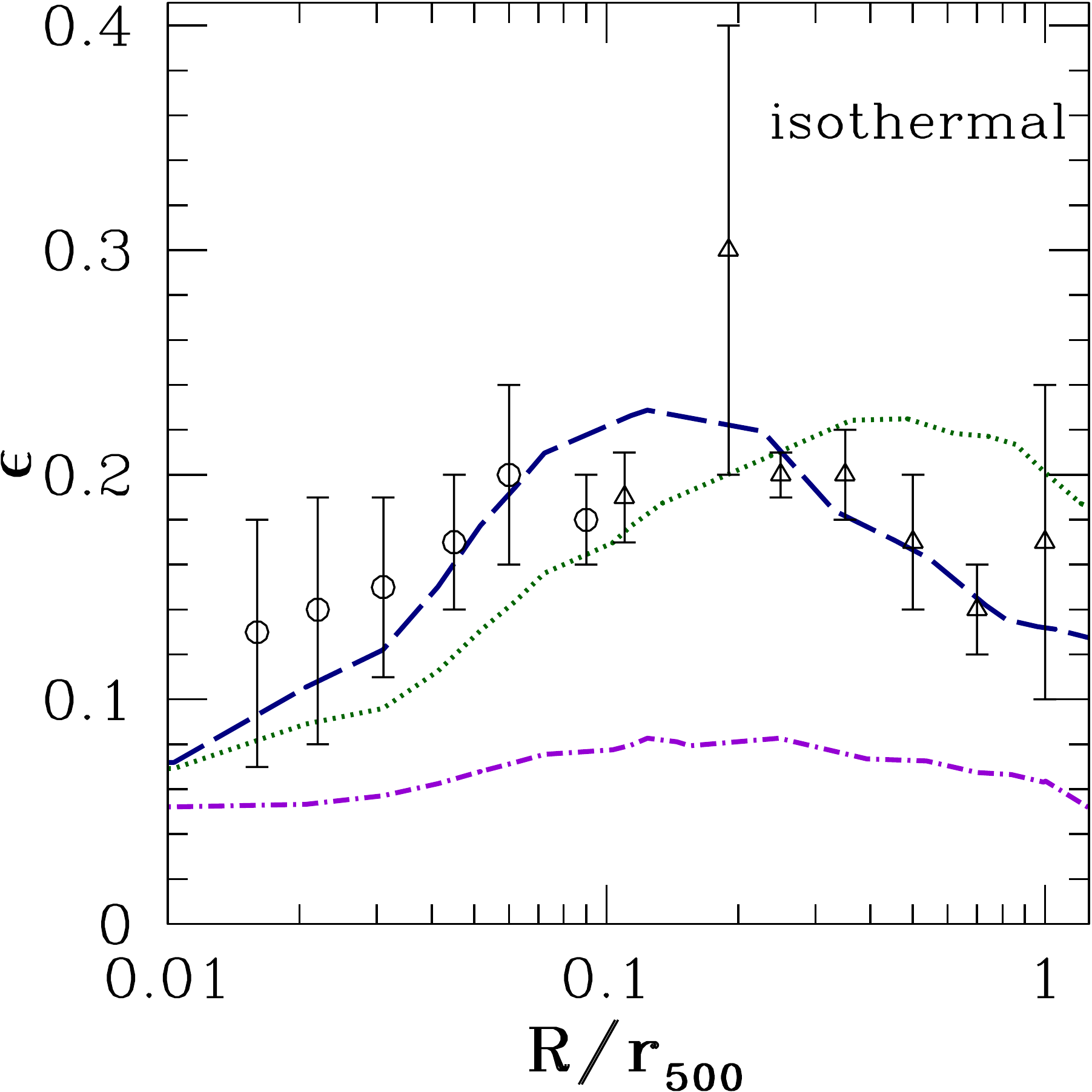}\includegraphics[width=0.5\linewidth, keepaspectratio]{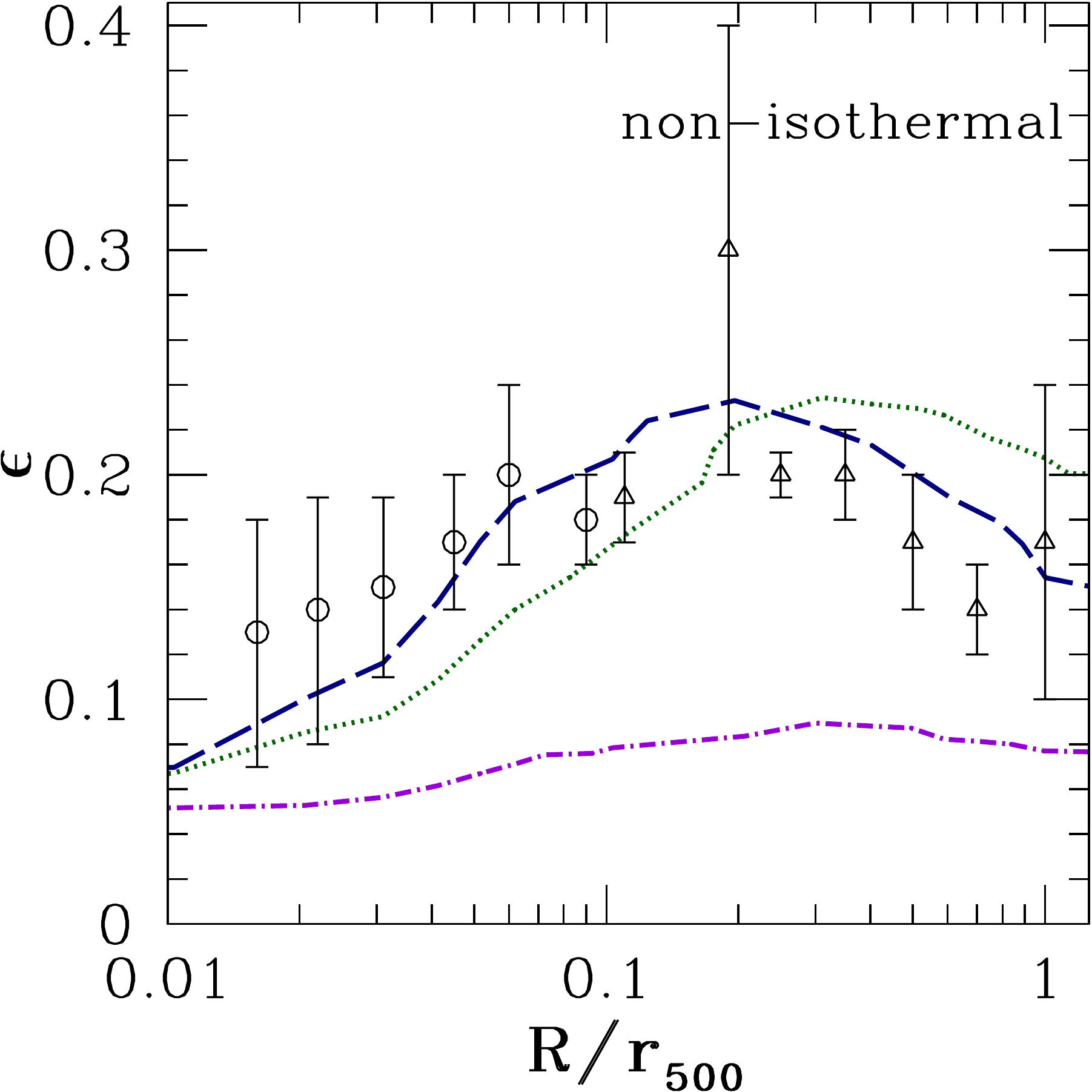}}
\caption{The
  ellipticity profiles of the isothermal and non-isothermal models (curves) compared with the average ellipticity profile of observed
  relaxed clusters (points): the circles show the ellipticity estimated using
  Chandra observations, while the triangles represent the measurements
  from ROSAT data \citet{lau12}. The left panel
  refers to the isothermal models, the right one to the non-isothermal
  models (I1 and NI1 models: dotted green line; I2 and NI2 models: dashed blue
  line; I3 and NI3 models: dot-dashed purple line). In all our models $r_{500}\approx1345\;\rm kpc$.}\label{lau12}
\end{figure*}

\subsection{X-ray ellipticity profiles}\label{500}

Thanks to the extensive coverage of X-ray surveys of galaxy
  clusters, the isophote ellipticity can be evaluated for a relatively
  large sample of observed objects. \citet{fang09} considered 10 relaxed
galaxy clusters from the ROSAT PSPC sample of \citet{buote96}, and
obtained a Chandra follow up for the central regions ($r\lesssim
100\rm\, kpc$) of the selected clusters. They evaluated an average ellipticity
profile with a constant value of $\epsilon\approx0.25$ up to $\approx
0.7\rm\, r_{500}$. This result is in agreement with
\citet{lau12}, who, using the low-redshift clusters sample from
\citet{vikhl09}, showed that the mean observed cluster ellipticity
is relatively constant with radius, with $\epsilon\approx0.18$ in the
radial range of $0.05\lesssim r/r_{500} \lesssim1$. In
  Fig.~\ref{lau12}, we show the ellipticity of our models superimposed
  to \citet{lau12} observations.

 We estimate the ellipticity $\epsilon$ of the isophotes in an
X-ray map following the same procedure as \citet{lau12}.  Our models
are consistent with the observational average ellipticity profile of
\citet{lau12}.  In particular, the I2 and NI2 models produce the
ellipticity profile with the best agreement with the observations. 
It is worth noting that the isothermal and non-isothermal
models produce similar ellipticity profiles when subjected to the same
velocity law. Nonetheless, the non-isothermal models are characterized by slightly higher values of
 $\epsilon$ in the cluster outskirts. The steeper decrease of
the rotation velocity along $R$ for the VP2 models with respect
to the VP1 models reflects into the steeper decrease of the ellipticity
profiles in the outer regions of the clusters.

\subsection{X-ray spectra}\label{sec:spec}
Here we discuss the effects of rotation on the X-ray emission
  lines of the model cluster spectra.  The measure of the centroid
position of the emission line profile is the most direct way of
probing the presence of ICM motions, but so far this method has been
mainly limited to theory (\citealt{ino03}, \citealt{rebusco08},
\citealt{zhura12}) due to the absence of an instrument with a
sufficiently high energy resolution. A first measure of the velocity
width of cool material in the X-ray luminous cores of a sample of
galaxy clusters, galaxy group and elliptical galaxies has been
obtained by \citet{sanders13}, who placed limits of the order of
$\lesssim300\,\rm km\, s^{-1}$ by fitting the emission line profiles
with a thermal model convolved with a Doppler broadening using the RGS
spectrometer on board the XMM telescope.

We build the simulated spectra using our ICM models. Thanks to the
high temperature, the emission line spectrum is characterised by the
presence of the Fe XXV ($\approx6.7\;\rm keV$) and Fe XXVI
($\approx6.9\;\rm keV$) emission lines. Their high emissivity makes
them particularly useful for the Doppler shift fitting procedure. We introduce the
effect of the gas rotation using the formalism of \citet{ino03}. The
authors consider that the classical Gaussian shape expected for an
emission line can be significantly shifted by large scale coherent gas motion and enlarged by turbulence.
Specifically, the Doppler shift can be expressed as
\begin{equation}
 \Delta E= 6.7(u_{\varphi}/300\;\rm [km\,s^{-1}])\;\rm{eV},
\end{equation}
where $u_{\varphi}$ is the ICM large scale velocity.
We simulate the observed spectra using the software XSPEC \citep{arnaud96}. First, we consider the case in which there is no turbulence, through the APEC model. Subsequently, we add a turbulent velocity component, using the BAPEC model.
\begin{figure}
 \begin{center}
   \includegraphics[width=1\linewidth,trim=0.9cm 0.0cm 0.5cm 0.45cm, clip=true, keepaspectratio]{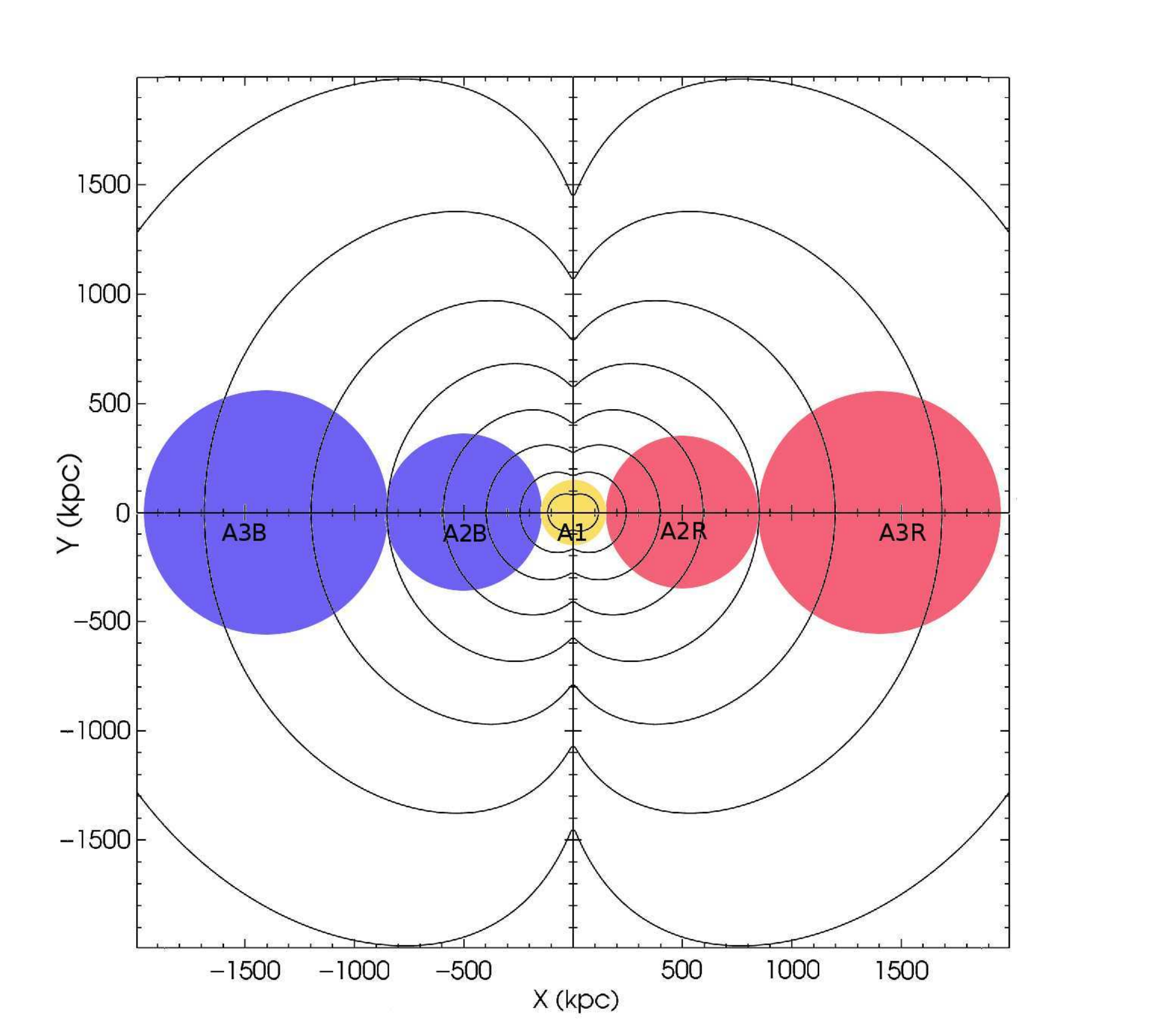}
\caption{The regions selected for the simulation of the spectra,
  plotted on the I1 model isophotes. The A1, A2, and A3 label
  indicate the centre of the inner, intermediate and external region,
  respectively. We distinguish the blue-shifted (approaching)
    regions (A2B, A3B) and red-shifted (receding) regions (A2R,
    A3R).}\label{zone}
 \end{center}
\end{figure}

\begin{figure}
 \begin{center}
   \includegraphics[width=0.7\linewidth,trim=0.0cm 1.1cm 1.5cm 0.0cm, clip=true, keepaspectratio, angle=90]{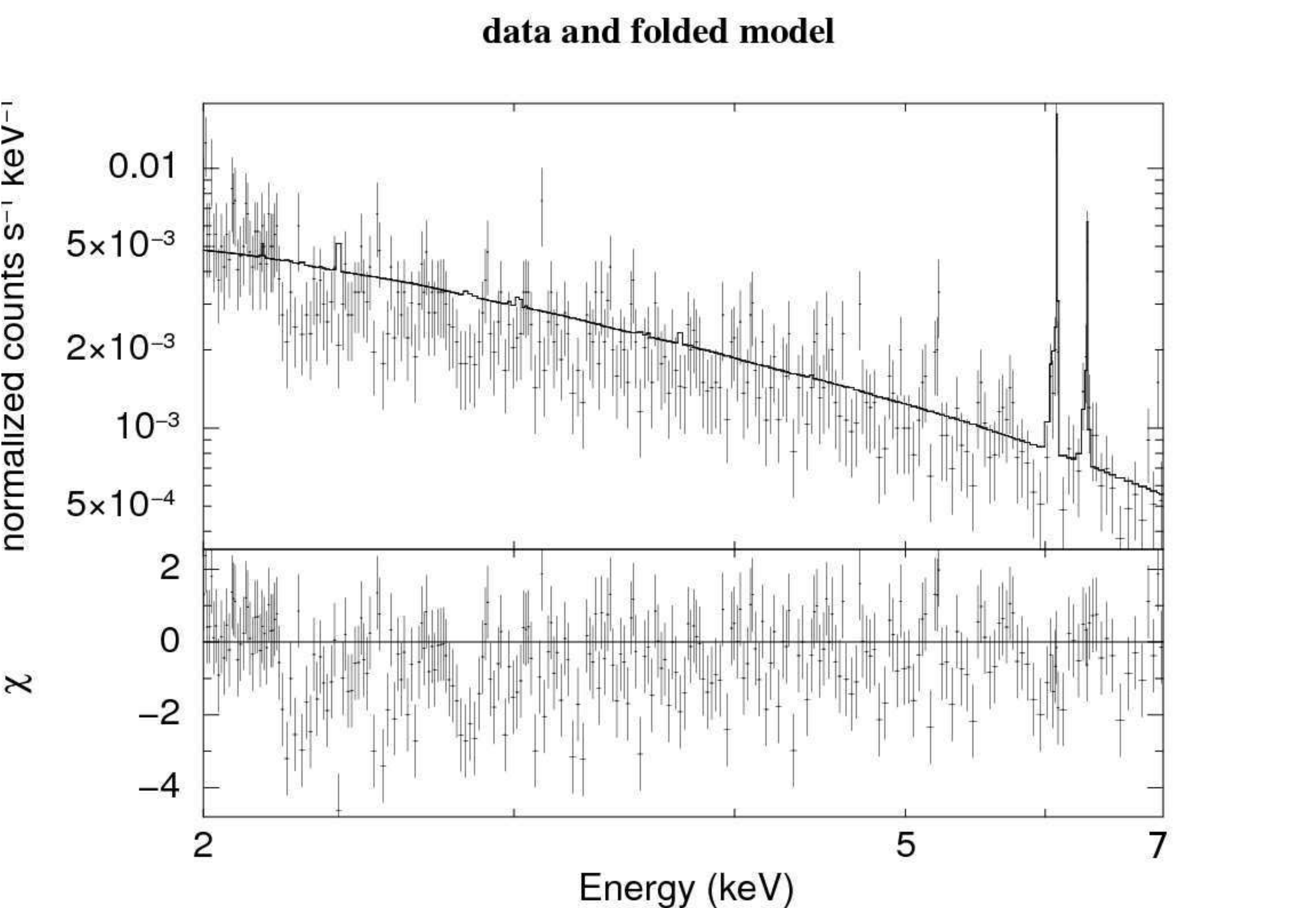}
\caption{The simulated spectrum of model I1 from the region A1. In the lower plot
  we show the residuals for the fitting with the APEC spectrum. The Fe
  XXV and Fe XXVI emission lines are particularly prominent around
  $E\approx6\;\rm keV$.  }\label{specstat}
 \end{center}

\end{figure}

 The cluster is assumed to be observed at redshift $z=0.1$. The
data are then convolved in XSPEC with an instrumental response
function of the X-ray Calorimeter Spectrometer on board the ASTRO-H
satellite, to create the final mock spectra. The convolution with the
instrument is performed without considering the background
contribution. We recall that the temperature distributions of the
models are fixed by choosing $T_{c}\simeq 7.5\rm \;keV$ as central
temperature (see Section~\ref{sec:fluideq}).  We consider the
simulated observations with an exposure time of $300$ ksec and
enabling the statistics model included in XSPEC. We select three
different circular regions A1, A2 and A3 with radii of $150 \; {\rm
  kpc} \; (1.4$ arcmin at the assumed redshift of 0.1), $350 \; {\rm
  kpc} \; (3.2 {\rm\; arcmin})$ and $550 \; {\rm kpc} \; (5 {\rm
  \;arcmin})$, respectively, and larger than the spatial extension of
about 0.5 arcmin of the single element array of the ASTRO-H Soft X-ray
Spectrometer System. Despite the symmetry of our models, we consider separately the
  blue-shifted (approaching) regions (A2B, A3B) and red-shifted
  (receding) regions (A2R, A3R) to account for the different
  response of the instrument at different energies (see Fig.~\ref{zone}).We divide the
regions A1, A2 and A3 into 20, 14, and 6 blocks respectively. We
assume a constant value of the density for each block, corresponding
to its central value. We projected the density onto the sky plane in
order to obtain the normalization constant for the models. Then, we
evaluate the line-of-sight component of the rotational velocity, being
this the factor responsible of the Doppler shift of the emission line
centre. In Fig.~\ref{specstat} we show the spectrum of model I1 obtained 
from the central region (A1): here, the low values of the rotational
velocity lead to a broadening of the emission line, and the line
centroid position remains unaffected.

\begin{figure*}
\centering
 \includegraphics[width=0.5\linewidth, keepaspectratio]{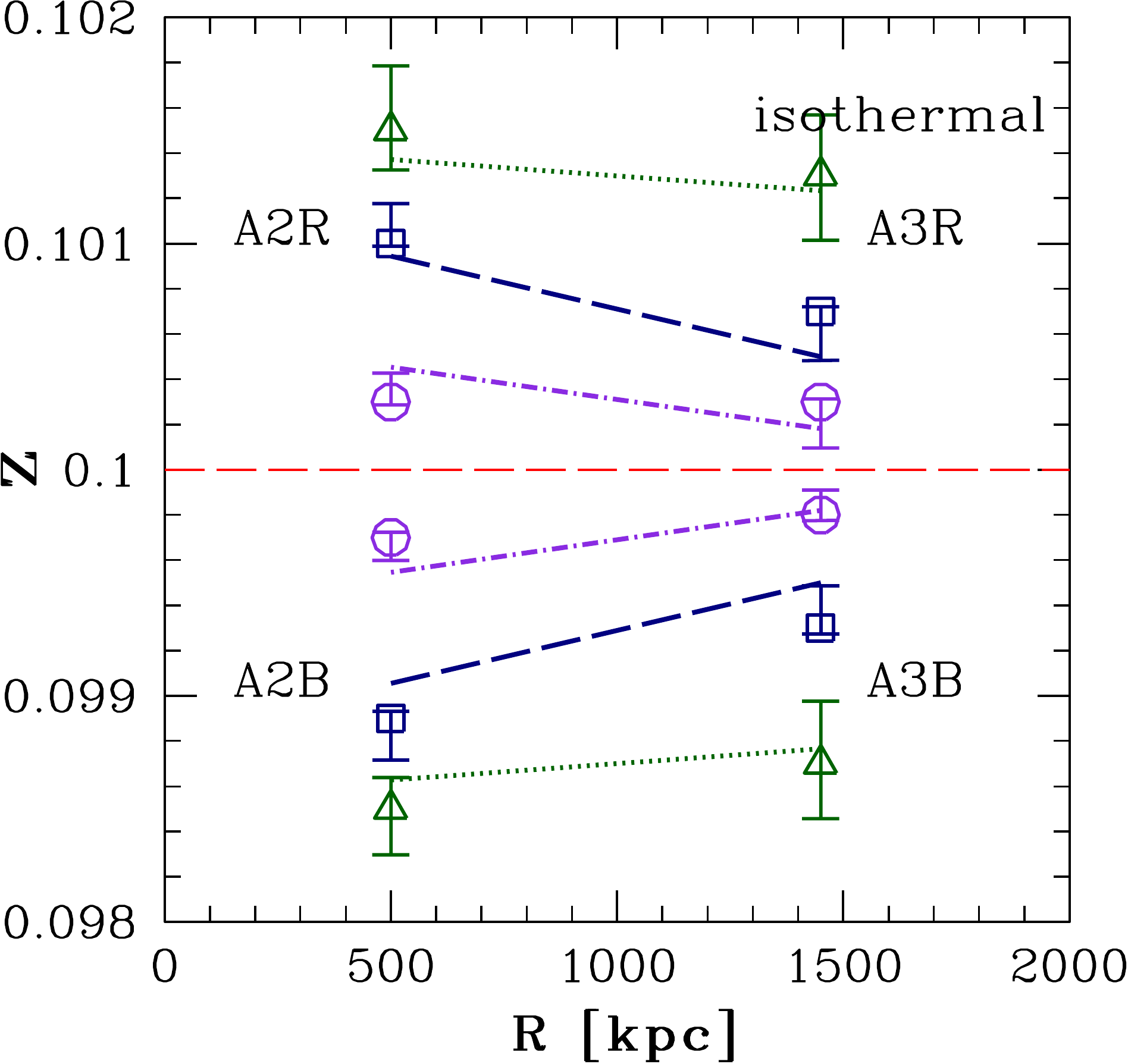}\includegraphics[width=0.5\linewidth, keepaspectratio]{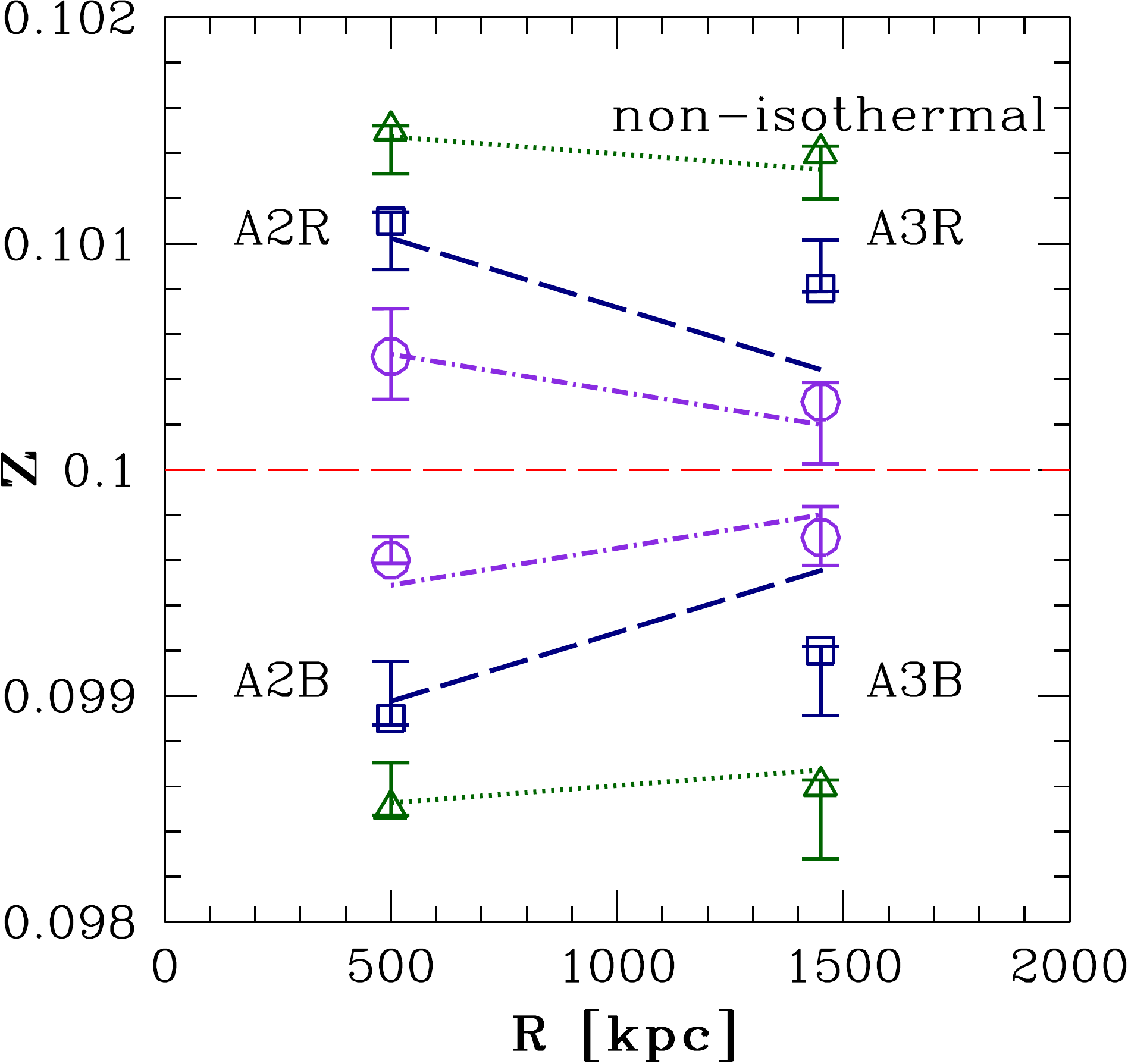}
\caption{ The Doppler shift fitting results for the spectra of
    models with no turbulence, obtained from the selected areas (A2B,
    A3B, A2R, A3R) as in Fig~\ref{zone}, plotted as a function of the
  distance from the cluster centre. The left plot refers to the
  isothermal models I1 (triangles), I2 (squares) and I3 (circles). The
  right plot refers to the non-isothermal models NI1 (triangles), NI2
  (squares) and NI3 (circles). The corresponding lines refer to the theoretical
  emission-weighted shift for each model. The horizontal dashed red line indicates the cluster redshift.}\label{blus}
\end{figure*}

For the rotating models, the resulting spectra from region A2 present
a shift ranging from $\approx10\;\rm eV$ for the VP1 model to $\approx6\;\rm
eV$ for the VP2 and $\approx5\;\rm eV$ VP3 models, in good agreement
with \citet{ino03} and with \citet{rebusco08}. We show the results of
the redshift fitting in Table~\ref{zfit} and in Fig.~\ref{blus},
plotted with the respective theoretical emission-weighted
shift. Figure~\ref{blus} shows that it is possible to discriminate 
between different models, both in the A2 and in the A3 centred
observations. It is worth noting that the red-shift and
  blue-shift results present a slight asymmetry (within the
  uncertainties) due to the different response of the instrument at
  different energies. The high central value of the VP2 velocity profile influences 
globally the fit of models,
 leading to fit values with the higher discrepancy with respect to the theoretical 
shift. We can also estimate the amplitude of the
detectable velocity constraining it to a lower limit of $\approx
100\;\rm km \,s^{-1}$. 


The inclusion of the turbulent velocity alters the centroid shift
fit. We use the parameters of the APEC models and added a turbulent
component $\sigma_{turb} = 200 \rm \,km\,s^{-1}$ through the BAPEC model included in
the XSPEC software. This value is consistent with the current observational limits
of $300-500\rm\; km\,s^{-1}$ for the inner regions of bright groups and clusters of galaxies 
\citep{sanders13}. 
 We report the results in Table~\ref{zfit}. The
Doppler shift fitting suffers from an higher error, but preserves a
good significance, in particular for the non-isothermal models.
We tested the fitting procedure up to $\sigma_{turb}=600\;\rm km\,^{-1}$, finding that 
the result remains significant ( $\sigma\approx 1$) up to $\sigma_{turb}=450\;\rm km\,^{-1}$
 We can
conclude that A2 is the most suitable region for the observation, in
which the higher gas density allows to obtain a better fit.

\begin{table*}
\centering
\begin{tabular} {| c   c  c c   c c}

\hline
\hline
Model & Region  &Observed shift &\multicolumn{2}{c}{{$\rm\sigma_{turb}=200 km\;s^{-1}$}}   \\

 &  &  $z_{\rm fit}(2\sigma)$, $\sigma$ &  $z_{\rm fit}(2\sigma)$, $\sigma$ & $\sigma_{turb,fit}$  \\

\hline
I1 & A2B & $0.0986^{+1.56\times10^{-4}}_{-3.43\times10^{-5}}$, $7.3$ & $0.0988^{+5.27\times10^{-4}}_{-5.47\times10^{-4}}$, $2.1$& $\leq304.24$ & \\
      & A2R  & $0.1016 ^{+ 2.76\times10^{-5}}_{- 1.45\times10^{-4}}$, $9.2$ & $0.1012 ^{+3.47\times10^{-4}}_{-3.08\times10^{-4}}$, $1.8$ & $\leq301.58$&\\
 & A3B &$0.0987^{+2.84\times10^{-4}}_{-2.51\times10^{-4}}$, $2.6$\\
 & A3R &$0.1013 ^{+ 2.12\times10^{-4}}_{-2.63\times10^{-4}}$, $2.7$\\
		      
\hline
I2 & A2B &$0.0989^{+3.25\times10^{-5}}_{-1.85\times10^{-4}}$, $5.0$&     $0.0988 ^{+3.45\times10^{-4}}_{-4.69\times10^{-4}}$, $1.4$& $214.25^{+105.7}_{-91.36}$ &\\
		     & A2R&$0.1010 ^{+ 1.77\times10^{-4}}_{-1.24\times10^{-5}}$, $5.2$ &$0.1011 ^{+1.87\times10^{-4}}_{-1.98\times10^{-4}}$, $2.8$ &$212^{+104.52}_{-94.57}$&\\
& A3B &$0.0993^{+1.88\times10^{-5}}_{-2.64\times10^{-4}}$, $2.5$ &\\
 &A3R&$0.1007 ^{+2.14\times10^{-5}}_{-2.17\times10^{-4}}$, $2.9$\\
\hline
I3 & A2B&$0.0997^{+2.39\times10^{-5}}_{-1.01\times10^{-4}}$, $2.5$&$0.0997 ^{+2.35\times10^{-4}}_{-1.07\times10^{-4}}$, $1.0$&$271.25^{+107.21}_{-109.71}$&\\
&A2R&$0.1003 ^{+1.28\times10^{-4}}_{-1.3\times10^{-5}}$, $2.2$&$0.1004^{+ 2.69\times10^{-4}}_{-1.25\times10^{-4}}$, $1.1$& $241.87^{+99.25}_{-112.91}$&\\
&A3B&$0.0998^{+1.09\times10^{-4}}_{-2.47\times10^{-5}}$, $1.5$\\
&A3R&$0.1003 ^{+2.03\times10^{-4}}_{-1.41\times10^{-5}}$ ,$1.3$\\
\hline
NI1 & A2B &$0.0985 ^{+ 2.24\times10^{-4}}_{-2.84\times10^{-5}}$, $5.9$&$0.0986^{+ 3.02\times10^{-4}}_{-2.82\times10^{-4}}$, $2.3$& $207.05^{+92.53}_{-87.58}$&\\
		     & A2R &$0.1015 ^{+ 2.14\times10^{-5}}_{-1.91\times10^{-4}}$, $7.1$& $0.1015^{+2.74\times10^-4}_{-3.03\times10^{-4}},  2.5$&$187^{+92.48}_{-87.56}$&\\
&A3B&$0.0986^{+2.78\times10^{-5}}_{-3.21\times10^{-4}}$, $4.0$\\
&A3R&$0.1014 ^{+3.12\times10^{-5}}_{-2.03\times10^{-4}}$, $2.7$\\
\hline
NI2 & A2B&$0.0989 ^{+ 2.53\times10^{-4}}_{-2.92\times10^{-5}}$, $3.8$&$0.0988 ^{+4.02\times10^{-4}}_{-2.58\times10^{-4}}$, $1.8$& $198.72^{+92.37}_{-81.48}$&\\
		     &  A2R &$0.1011 ^{+ 3.87\times10^{-5}}_{-2.15\times10^{-4}}$, $4.3$&$0.1011 ^{+4.26\times10^{-4}}_{-3.25\times10^{-4}}$, $1.4$& $174^{+94.5}_{-101.84}$& \\
&A3B&$0.0992^{+1.88\times10^{-5}}_{-2.87\times10^{-4}}$, $2.6$\\

&A3R&$0.1008 ^{+2.15\times10^{-4}}_{-1.23\times10^{-5}}$, $3.5$\\
\hline
NI3 &  A2B  &$0.0996^{+1.05\times10^{-4}}_{-1.48\times10^{-5}}$, $3.4$&$0.0997 ^{+1.14\times10^{-4}}_{-1.35\times10^{-4}}$, $1.2$& $208.25^{+89.54}_{-92.36}$&\\
	& A2R& $0.1005^{+2.11\times10^{-4}}_{-1.89\times10^{-5}}$, $2.2$&$0.1005^{+1.38\times10^{-4}}_{-3.11\times10^{-4}}$, $1.2$& $175.36^{+71.69}_{-85.36}$&\\
&A3B&$0.0997^{+1.38\times10^{-4}}_{-1.23\times10^{-4}}$, $1.2$\\
&A3R&$0.1003^{+8.65\times10^{-5}}_{-2.74\times10^{-4}}$, $2.6$ \\
\hline
\hline

CC1 & A2B &$0.0986^{+2.23\times10^{-4}}_{-3.11\times10^{-5}}$, $5.5$&
	  $0.0986^{+ 4.27\times10^{-5}}_{-2.89\times10^{-4}}$, $4.2$&  $221.3^{+89.4}_{-73.5}$&\\
	& A2R &$0.1014 ^{+ 1.98\times10^{-4}}_{-3.02\times10^{-5}}$, $6.1$&
	      $0.1013^{+3.25\times10^{-5}}_{-3.62\times10^{-4}}, 3.2$&$198.3^{+86.2}_{-90.2}$&\\
&A3B&$0.0987^{+ 3.84\times10^{-4}}_{-2.54\times10^{-4}}$, $2.1$\\
&A3R&$0.1013 ^{+2.01\times10^{-4}}_{-2.36\times10^{-4}}$, $2.9$\\
\hline
CC2 & A2B&$0.0988 ^{+ 2.18\times10^{-4}}_{-1.35\times10^{-5}}$, $5.1$&
	  $0.0989 ^{+1.74\times10^{-4}}_{-3.98\times10^{-4}}$, $1.9$& $206.7^{+89.5}_{-92.5}$&\\
		     &  A2R &$0.1012 ^{+ 8.57\times10^{-5}}_{-1.58\times10^{-4}}$, $5.0$&
		  $0.1008 ^{+1.56\times10^{-4}}_{-2.24\times10^{-4}}$, $2.1$& $196.8^{+102.5}_{-84.9}$& \\
&A3B&$0.0993^{+3.02\times10^{-4}}_{-2.48\times10^{-5}}$, $2.2$\\

&A3R&$0.1005 ^{+2.11\times10^{-4}}_{-1.71\times10^{-5}}$, $2.1$\\
\hline
CC3 &  A2B  &$0.0994^{+1.85\times10^{-4}}_{-2.71\times10^{-5}}$, $2.8$&
	  $0.0994 ^{+3.89\times10^{-4}}_{-1.18\times10^{-4}}$, $1.2$& $215.9^{+98.1}_{-105.8}$&\\
	& A2R& $0.1004^{+3.15\times10^{-5}}_{-1.54\times10^{-4}}$, $2.2$&
	  $0.1004^{+1.54\times10^{-4}}_{-2.89\times10^{-4}}$, $1.1$& $199.8^{+102.5}_{-95.4}$&\\
&A3B&$0.0997^{+3.55\times10^{-5}}_{-1.12\times10^{-4}}$, $2.0$\\
&A3R&$0.1003^{+2.84\times10^{-5}}_{-1.25\times10^{-4}}$, $1.9$ \\
\hline
  \end{tabular}
\caption{ Spectral constraints on the rotating ICM. From left to
  right, we have the model name (column 1), the region considered
  (column 2), the result of the Doppler shift fitting of the rotating
  models and its significance (column 3), and the Doppler shift
  fitting of the rotating models with a turbulent component and its
  significance (column 4). We recall that when computing the 
mock spectra we assume that the clusters are at redshift $z=0.1$.} \label{zfit}
\end{table*}

\section{Effect of cool cores}\label{sec:cool}

\begin{figure}
 \begin{center}
   \includegraphics[width=0.9\linewidth]{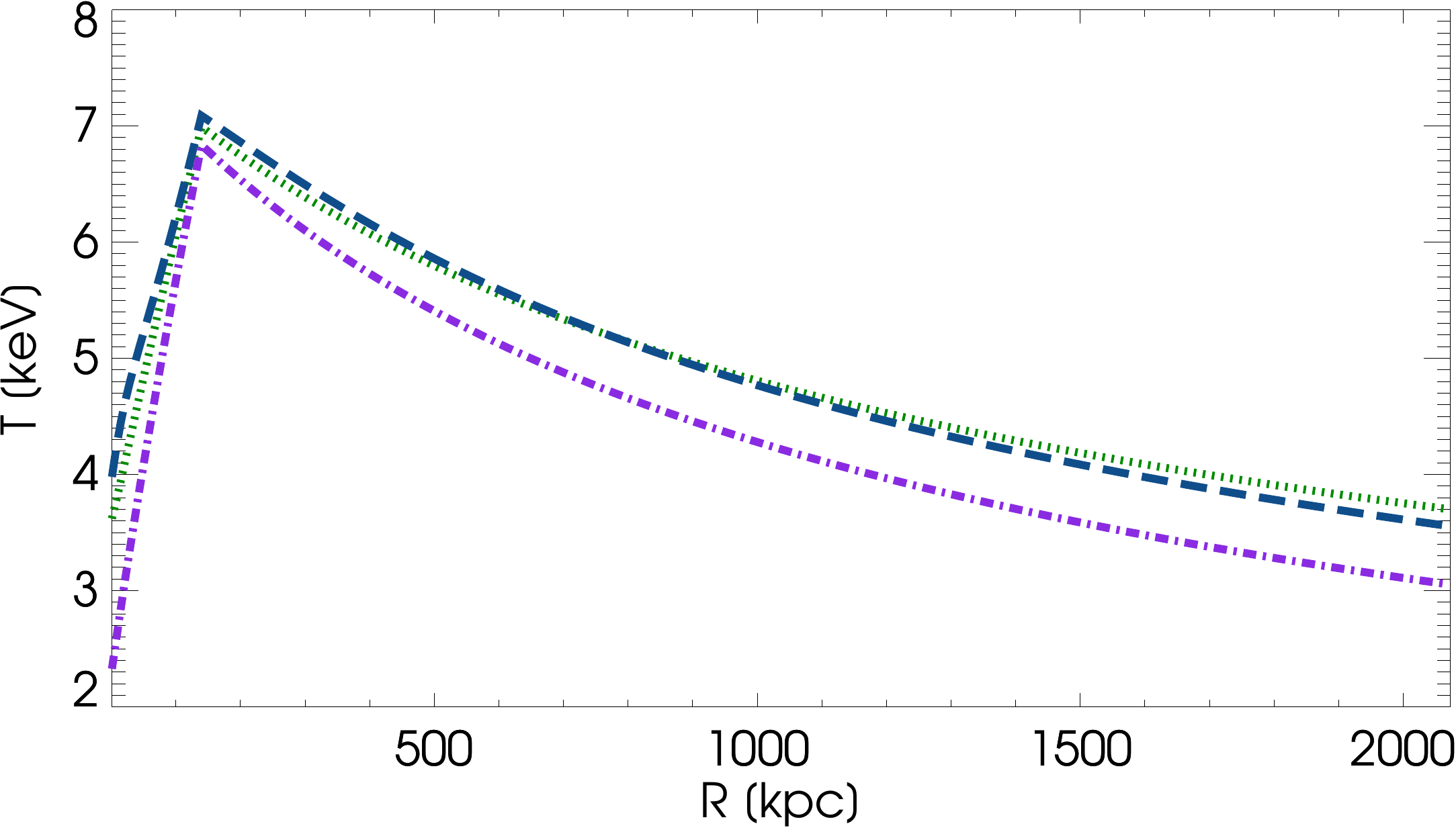}
\caption{The temperature profiles along $z=0$ for the CC models CC1 (dotted green line),  
CC2 (dashed blu line)
  and  CC3 (dot-dashed purple line). }\label{tempcc}
 \end{center}
\end{figure}

So far we have considered models of non-cool core clusters.  In this
section, we discuss the effect of cool cores on the ICM properties
investigated in our study.  Here we limit ourselves to presenting toy
models of rotating clusters with cool cores, assuming that these
systems are stationary, and so neglecting the complex interplay of
cooling and heating in the central regions. In this spirit, we assume
that these rotating cool-core (CC) models have the same temperature
profiles in the outer regions as the NI models, but we impose an
outward increasing temperture profile in the core.
To reproduce this behaviour, we use composite polytropes, assuming the following 
relation between pressure and density: 
\vspace{2cm}
\begin{equation}
\dfrac{P}{P_0}=\left(\dfrac{\rho}{\rho_0}\right)^{\tilde\gamma_{in}} \rm \;\;\;\ if \;\rho>\rho_0,
\end{equation}
\begin{equation}
\dfrac{P}{P_0}=\left(\dfrac{\rho}{\rho_0}\right)^{\tilde\gamma_{out}} \rm \;\;\;\ if \; \rho<\rho_0,
\end{equation}
where $\rho_0$ and $P_0$ are the gas density and pressure at the boundary 
between the inner (core) region and the outer region, and we assume 
$\tilde\gamma_{in}=0.59$ and $\tilde\gamma_{out}=1.14$. We present 
here three CC models, named CC1, CC2, CC3, having respectively velocity
 profiles VP1, VP2, VP3. The global properties of these models are the 
same as for models I and NI (see Table~\ref{common}). 
In terms of the gas number density $n_0$ and temperature $T_0$ at the boundary
 (related to pressure and density by  $P_0=n_0k_BT_0$ and $\rho_0=\mu m_p n_0$, 
where $\mu=0.59$ 
and $m_p$ is the proton mass), we assume  $n_0= 5.37\, \rm\times 10^{-3}\, cm^{-3}$ , $k_BT_0=6.9\, \rm keV$ for 
model CC1, $n_0=6.31 \, \rm\times 10^{-3} cm^{-3}$ , $k_BT_0=7.1\, \rm keV$ 
for model CC2, and $n_0=11.48\, \times10^{-3}  \rm cm^{-3}
$, $k_BT_0=6.79 \, \rm keV$ for model CC3.
In Fig.~\ref{tempcc} we plot the temperature profiles in the $z=0$ plane, for the rotating CC models.

\begin{figure*}
 \begin{center}
   \includegraphics[width=0.5\linewidth]{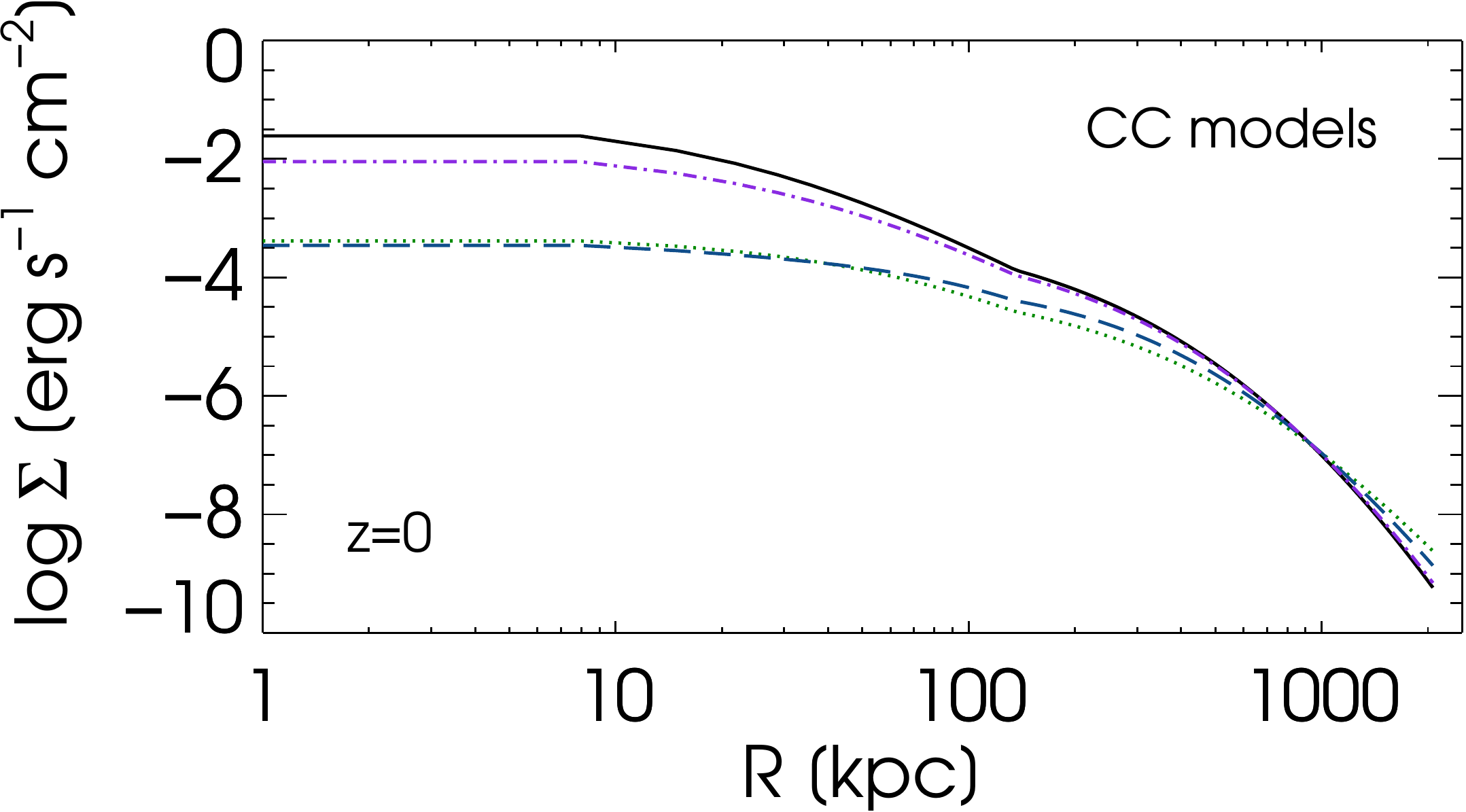}\includegraphics[width=0.5\linewidth]{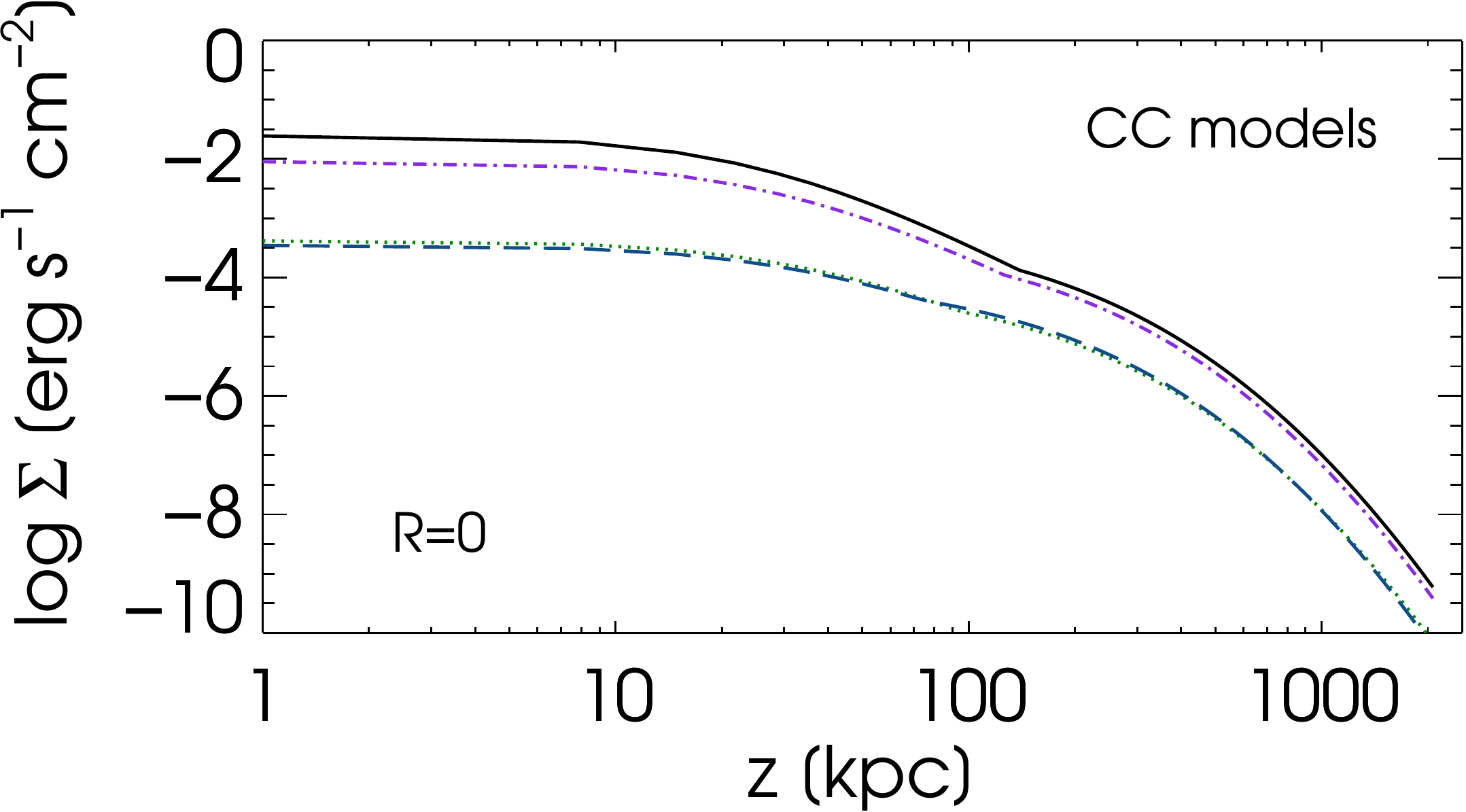}
\caption{ The surface brightness profile along $z=0$ (left panel) and
  $R=0$ (right panel), for the cool core models. Specifically, the CC1
   (dotted green line), CC2 (dashed blu line), and CC3 (dot-dashed purple line) models are shown.
   The solid lines represent the corresponding non-rotating model}\label{britcc}
 \end{center}
\end{figure*}

\begin{figure}
 \begin{center}
   \includegraphics[width=\linewidth]{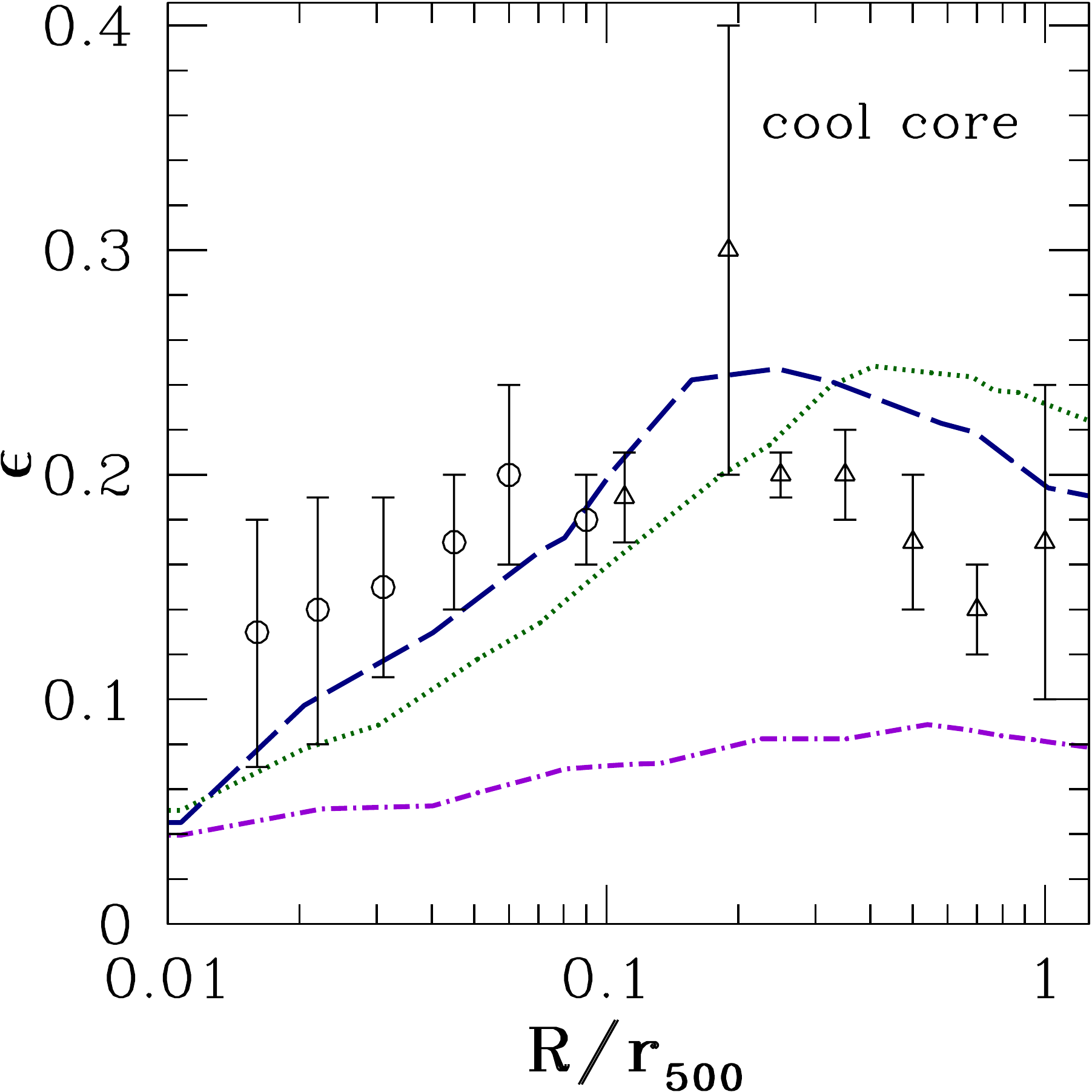}
\caption{The
  ellipticity profiles of our CC models (curves). The
different line types represent the cool core models as in Fig.~\ref{tempcc}. 
The symbols with error bars are the same as in Fig.~\ref{lau12}.}\label{ccmod}
 \end{center}
\end{figure}

\begin{figure}
 \begin{center}
   \includegraphics[width=\linewidth]{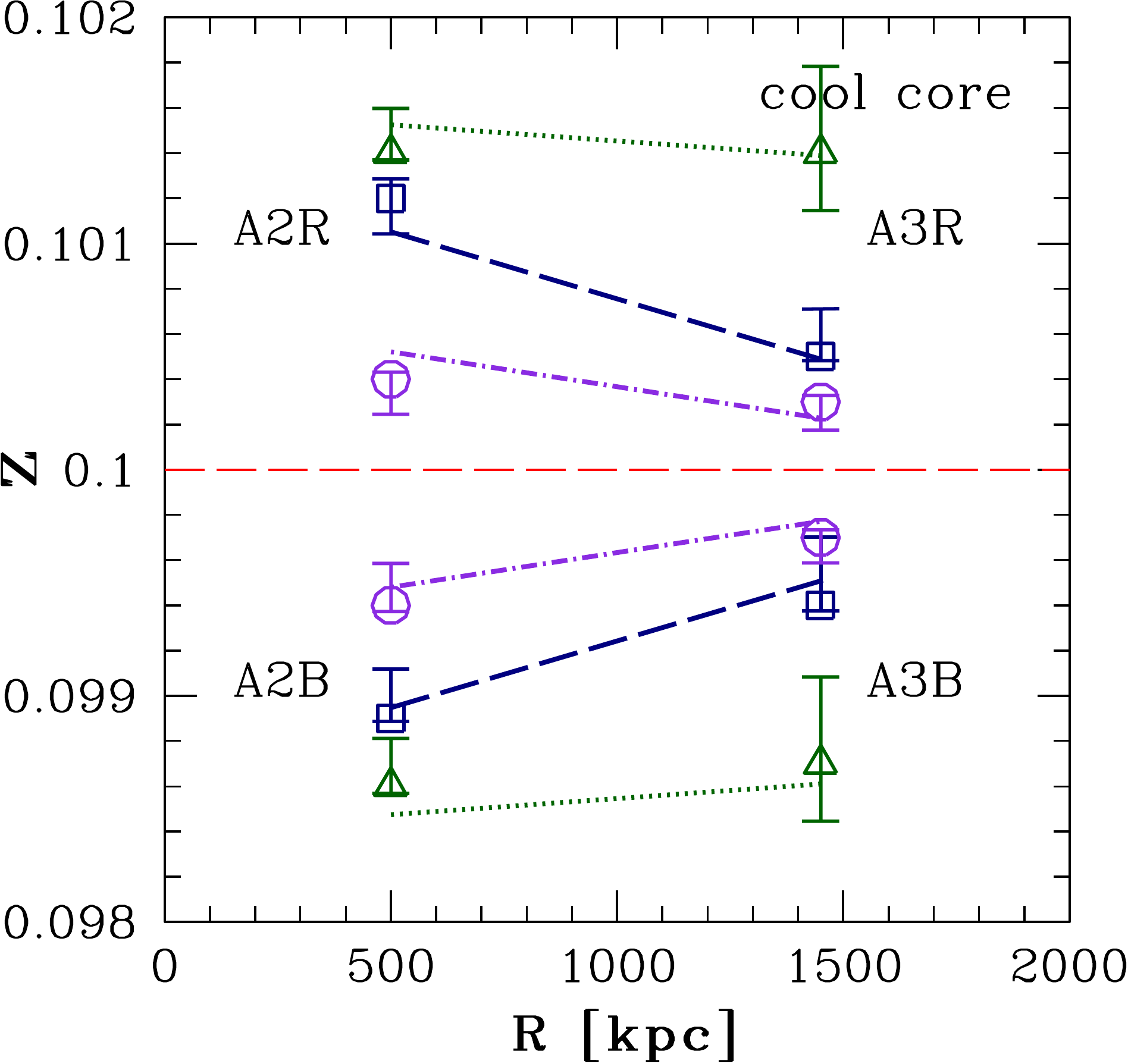}
\caption{Same as Fig.~\ref{blus}, but for
  the cool core models CC1 (triangles), CC2 (squares) 
and CC3 (circles).}\label{ccpeso}
 \end{center}
\end{figure}

We obtain the surface brightness profiles along the $z=0$ and $R=0$ axes, shown
in Fig.~\ref{britcc}, where we included the composite non-rotating model, 
having the same gas fraction as the corresponding rotating models.
 The surface brightness profiles of the rotating models along $R=0$ and $z=0$ present a 
depletion of the inner regions comparable to the I and NI models.
As expected, the CC models have higher central surface brightness compared to the NI models
and steeper inner surface brightness profiles. The ellipticity profiles of the CC models 
are shown in Fig~\ref{ccmod}: the profiles are similar to those of the I and NI models, and therefore 
 consistent with the observations by \cite{lau12}.
Following the procedure in Section~\ref{sec:spec}, we build the simulated spectra of
the CC models. The resulting spectra from region A2 present
a shift ranging from $\approx11\;\rm eV$ for the VP1 model to $\approx6\;\rm
eV$ for the VP2 and $\approx5\;\rm eV$ VP3 models, similar to the values
obtained for the non-cool core models (see Section~\ref{sec:spec}). We show the results of
the redshift fitting in Table~\ref{zfit} and in Fig.~\ref{ccpeso},
plotted with the corresponding theoretical emission-weighted
shift. We note again a slight asymmetry (within the
  uncertainties) due to the different response of the instrument at
  different energies. Globally, the CC models maintain the same properties as the I and NI models.

\section{Conclusion}
The study of the dynamics of the intracluster medium is fundamental to
depict the physical processes involved in the cluster evolution. In
this work, we demonstrate the capability of the X-ray observations to
characterise the features induced from a rotating ICM on the shape of
the surface brightness isophotes and on the spectral emission lines. 
This can be extended to test the hydrostatic equilibrium
hypothesis on which the majority of the current X-ray mass estimate and
calibrations of the scaling laws relies. We consider, as case study, representative models
of massive galaxy clusters with a spherical dark matter halo. A simple
modelling of the gas, obtained through the assumption of a polytropic
distribution $P\propto\rho^{\tilde{\gamma}}$, allow us to span the
whole range of the observed ellipticity profiles and to reproduce the
observed values of \citet{lau12}. For instance, we find good agreement with the 
observed ellipticities for the rotation profile VP2, with a peak velocity 
of $\approx 600\rm\;km \,s^{-1}$ at $0.05 \;\rm r_{200}$ decreasing to $\approx 130\rm\;km \,s^{-1}$ at $\rm r_{200}$.
 We demonstrate also the capability
of the future generation of the X-ray calorimeter expected to fly
onboard the next generation of X-ray satellites, like the X-ray
Calorimeter Spectrometer onboard the ASTRO-H, to investigate some of
the features induced by rotating ICM. Thanks to its excellent energy
resolution (with a goal of $\Delta E\approx4 \;\rm eV$), such a
calorimeter is suitable for detecting the Doppler shift ($\Delta
E\approx 10 \;\rm eV$, $\Delta E\approx 6 \;\rm eV$ and $\Delta
E\approx5 \;\rm eV$ for our VP1, VP2 and VP3 models, respectively) and
the emission line broadening (of the order of $\approx100\rm\, km\;
s^{-1}$) on the ICM emission lines with a good significance. Together
with a resolved imaging of the ICM isophotes and the determination of
their ellipticity, a calorimeter-like detector will permit to
discriminate among the different velocity profiles here
considered. We have extended this procedure to cool-core clusters, 
by using composite polytropic profiles. 
Overall, the cool-core models are similar, 
in terms of ellipticity profile and spectral line shift, to the corresponding non cool-core models.
 A natural extension of our work would be the
  introduction of less crude rotation velocity patterns, for instance
  considering baroclinic distributions, in which the azimuthal
  velocity is not stratified on cylinders. This would help in
obtaining a more regular gas morphology. Also the question of the
  thermal stability of these rotating models has to be addressed in
  detail. Finally, it would be useful to compare the ICM shape
obtained with high-resolution SZ imaging. This is a potentially
powerful means of extending this analysis to higher redshifts, because
of the redshift independence of the SZ effect.

\section*{Acknowledgements}
The authors thank the anonymous referee for useful comments that
improved the presentation of the work.
MB acknowledges the financial support by the Austrian
Science Fund (FWF) through grant P23946-N16. MB thanks Andrea Negri and Dominik Steinhauser for helpful discussions.
SE acknowledges the financial contribution from contracts ASI-INAF I/023/05/0 and I/088/06/0.
CN acknowledges financial support from PRIN MIUR 2010-2011, project ``The 
Chemical and Dynamical Evolution of the Milky Way and Local Group 
Galaxies'', prot. 2010LY5N2T.

\end{document}